\documentclass[journal,onecolumn,12pt]{IEEEtran}
\usepackage{cite}
\usepackage{graphicx}
\usepackage{subfigure}
\usepackage{mathrsfs}
\usepackage{amsmath}
\usepackage{amsfonts}
\usepackage{amssymb}
\usepackage{fancybox}

\allowdisplaybreaks

\newcommand{\qed}{\hspace*{\fill} $\Box$ \\}
\newcommand{\bs}{\boldsymbol}

\newcommand{\bi}{\begin{itemize}}
\newcommand{\ei}{\end{itemize}}

\def\ba{\begin{array}}
\def\ea{\end{array}}
\def\be{\begin{equation}}
\def\ee{\end{equation}}
\def\bea{\begin{eqnarray}}
\def\eea{\end{eqnarray}}
\def\beas{\begin{eqnarray*}}
\def\eeas{\end{eqnarray*}}

\newtheorem{theorem}{Theorem}
\newtheorem{corollary}{Corollary}
\newtheorem{definition}{Definition}
\newtheorem{lemma}{Lemma}

\title{Pricing, Competition, and Routing for Selfish and Strategic
Nodes in Multi-hop Relay Networks}
\author{\authorblockN{Yufang Xi and Edmund M. Yeh}\\
\authorblockA{Department of Electrical Engineering\\
Yale University\\
New Haven, CT 06520, USA \\ Email: \{yufang.xi, edmund.yeh\}@yale.edu}}

\begin{document}

\maketitle

\begin{abstract}

We study a pricing game in multi-hop relay networks where nodes
price their services and route their traffic selfishly and
strategically. In this game, each node (1) announces pricing
functions which specify the payments it demands from its respective
customers depending on the amount of traffic they route to it and
(2) allocates the total traffic it receives to its service
providers. The profit of a node is the difference between the
revenue earned from servicing others and the cost of using others'
services. We show that the socially optimal routing of such a game
can always be induced by an equilibrium where no node can increase
its profit by unilaterally changing its pricing functions or routing
decision. On the other hand, there may also exist inefficient
equilibria. We characterize the loss of efficiency by deriving the
price of anarchy at inefficient equilibria.  We show that the price
of anarchy is finite for oligopolies with concave marginal cost
functions, while it is infinite for general topologies and cost
functions.

\end{abstract}

\section{Introduction}\label{sec:Intro}

It has been widely recognized that cooperation in networks formed by autonomous and
selfish nodes cannot be achieved unless sufficient incentives are provided to the nodes.
Such incentives normally take the form of payment or reward to the nodes if they help
forward other nodes'
traffic~\cite{paper:IMM05,paper:BLV05,paper:CGKO04,paper:MQ05,book:BH07}. A node is
usually willing to participate in routing only if it can charge more than the cost of
servicing the transit traffic. While a selfish node always prices its service with the
ultimate aim of maximizing its profit, it has to do so strategically since the customers
it courts may potentially buy services from other nodes. Thus, there exists a trade-off
in each node's pricing decision. That is, higher charges yield larger profit margins but
risk losing market share to its competitors.

In this work, we study the game that arises from the selfish and
strategic pricing behavior of relay nodes in a unicast multi-hop
relay network consisting of one source and one destination. A node
is selfish in the sense that maximizing its own profit is its sole
objective. Being strategic means that a node is able to optimally
design its pricing based on the anticipation of its competitors'
best response to its action. Specifically, in this game each node is
a service provider to a group of nodes (its customers), and when it
needs to forward the traffic received from its customers, the node
itself becomes a customer that uses the services of some other group
of nodes. As a service provider, the node announces pricing
functions which specify the payments it demands from its respective
customers depending on the amount of traffic the customers route to
the node. As a customer, the node allocates the total traffic it
receives to its service providers in a way that minimizes the sum of
its own transmission costs and the payments made to the service
providers. Such a game can exist in both wireline and wireless
networks, where communications consume resources and nodes are often
selfish agents. When a network, especially a wireless network, is
formed in an ad hoc manner, a node is typically aware of its
neighbors only. A rational node thus always bases its pricing and
routing decisions on the strategies adopted by its neighbors. We
will show that such a game always has equilibria where no node can
increase its profit by unilaterally changing its pricing functions
or routing decision. Furthermore, depending on the network topology
and the nodes' response strategy, the global routing configuration
at an equilibrium may or may not be socially efficient. We
characterize the loss of efficiency by deriving the price of anarchy
at inefficient equilibria. It is found that the price of anarchy is
finite for some link cost functions and topologies, while it is
infinite for others.

Pricing schemes were introduced into network resource allocation problems first as a means of
decomposing a global optimization into sub-problems solved by individual agents~\cite{paper:KMT98}.
In addition to being a facilitating device, pricing serves as an
essential mechanism for inducing social efficiency when users
(source nodes) selfishly choose their routes~\cite{paper:CDR03}. It
is well known that without appropriate pricing, e.g. marginal cost
pricing, selfish routing inevitably results in loss of efficiency,
which in general can be arbitrarily
large~\cite{paper:Rou02,paper:Rou05}. 

When service providers are also mindful of their self interest, they
will use pricing to their own advantage rather than to heed any
social mission. With both users and service providers behaving
selfishly, the network increasingly approximates a free market,
where prices can assume a variety of functions and lead to direct or
indirect competition among service providers. For example, pricing
network services according to their quality helps to match each type
of service with the customers that value it the
most~\cite{paper:SV03,paper:HW05}. By modelling the interaction
between the service provider and the users as a Stackelberg game,
\cite{paper:BS02} shows that when the service provider always adopts
the profit-maximizing price, its revenue per unit bandwidth and the
net utility of each user both improve with the number of users. When
multiple service providers are present in a network, price
competition inevitably
ensues~\cite{paper:HeW05,paper:SS05,paper:AO06}. It is demonstrated
in~\cite{paper:HeW05,paper:SS05} that cooperation in pricing is in
the best interest of service providers who jointly serve the same
customers. The dire consequence of non-cooperation is explicitly
analyzed in~\cite{paper:AO06}, which shows that price competition in
parallel-serial networks can result in arbitrarily large efficiency
loss. 

In this paper, we analyze the pricing game in multi-hop relay
networks where a node can compete for traffic from multiple nodes
and can allocate its received traffic to multiple nodes. Thus, in
general, a node is both a service provider and a customer. Another
distinctive feature of the game we consider is that the bid from
each service provider to a targeted customer is a (possibly
nonlinear) \emph{pricing function}, which specifies the price
contingent on the amount of service provided. Previous work on
pricing games almost exclusively assume a constant unit price from
every service provider, which in our terms means restricting pricing
functions to be linear. It turns out that the generalization from
linear to nonlinear pricing allows for a much richer set of
possibilities in pricing games. Even in economics literature, the
issue of nonlinear pricing is quite new and
challenging~\cite{book:Wil93}.\footnote{The nonlinear pricing game
we study can be seen as a generalized menu auction~\cite{paper:BW86}
where each bidder offers a continuum of options along with their
prices.}
Equilibria derived from such a general framework represent the most
fundamental outcomes of pricing games in multi-hop networks.

We show that the socially optimal routing can always be induced by
an equilibrium of the routing/pricing game where no node can
increase its profit by unilaterally changing its pricing functions
or routing decision. On the other hand, there also exist inefficient
equilibria. In particular, we show that in an oligopoly
routing/pricing game, inefficient equilibria are always
monopolistic, i.e., a dominant relay carries all the flow from the
source. We prove that the price of anarchy at such inefficient
equilibria is equal to the number of relays in an oligopoly if
marginal cost functions are concave. In this case, the worst
inefficient equilibria arise with linear marginal cost functions.
When marginal cost functions are convex, however, the price of
anarchy can be arbitrarily large.  Unlike the case of oligopolies,
inefficiency in general multi-hop relay networks stems not only from
dominant relays exercising monopolistic pricing power, but also from
the myopia of dominant relays. We demonstrate that the inability of
a node to gauge the impact of its pricing beyond its local
neighborhood can lead to an infinitely large price of anarchy.

\section{Network Model and Problem Formulation}\label{sec:Model}

\subsection{Network Traffic and Multi-hop Routing}\label{sec:network}

We consider a relay network represented by a directed graph
$\mathcal G = ( {\cal N}, {\cal E} )$ with one \emph{source} $s$ and
one \emph{destination} $w$, and a set of \emph{relays} which can be
used to forward traffic in a multi-hop fashion from $s$ to $w$. The
source $s$ needs to send traffic of a fixed rate $R_s$ to
$w$,\footnote{We will discuss the problem involving an elastic
session in Section~\ref{sec:ElasticPG}.} which can be carried
through links in ${\cal E}$.

We assume that there is no direct link between $s$ and $w$. That is,
traffic from $s$ has to be routed to $w$ via relays in a multi-hop
fashion. To make matters simple, we assume ${\cal G}$ contains only
nodes and links which are on the paths from $s$ to $w$. Since route
discovery is not a main concern of this work, we assume such a
${\cal G}$ is given a priori, and is loop-free. Each node, however,
needs only to be aware of its neighbors (predecessors, siblings, and
offsprings) as specified below.

\begin{figure}[!h]
  \begin{center}
  \includegraphics[width=5cm]{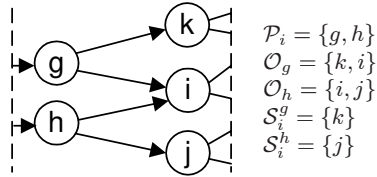}\\
  \caption{Illustration of predecessors, siblings and offsprings.}\label{fig:Notation}
  \end{center}
\end{figure}

For node $i$, $h$ is a \emph{predecessor} if $(h,i)\in {\cal E}$.
Denote the set of $i$'s predecessors by ${\cal P}_i$. For any $h \in
{\cal P}_i$, define ${\cal S}_i^h \triangleq \{j \ne i: h \in {\cal
P}_j\}$. That is, ${\cal S}_i^h$ is the set of nodes which share the
common predecessor $h$ with $i$. These are the nodes who compete
with $i$ for $h$'s traffic in the pricing game to be introduced
later. We will refer to them as \emph{siblings} of $i$ with respect
to $h$. Finally, $i$ is said to be an \emph{offspring} of $h$ if
$(h,i) \in {\cal E}$. Let the set of $h$'s offsprings be denoted by
${\cal O}_h$. The above notation is illustrated in
Figure~\ref{fig:Notation}.
We make a simplifying yet plausible assumption that $\bigcup_{h \in {\cal P}_i}{\cal S}_i^h \cap {\cal P}_i =
\emptyset$, i.e., no node can be both a sibling and a predecessor of any other node.

By our assumption on ${\cal G}$, $s$ is the only node without any predecessor while $w$
is the only node without any offspring. Since the pricing game to be studied can arise
only if there are multiple relays competing for the traffic from their common
predecessor, we assume in $\cal G$ that every node $i$ except $w$ has multiple relays in
${\cal O}_i$ unless ${\cal O}_i = \{w\}$. 

Denote the rate of flow on $(i,j) \in {\cal E}$ by $f_{ij}$. A link
flow vector $\bs f \triangleq (f_{ij})_{(i,j) \in {\cal E}}$ is a
\emph{routing} of the session traffic if it satisfies the flow
conservation constraint: $\sum_{h \in {\cal O}_s} f_{sh} = R_s$,
$\sum_{k \in {\cal P}_w} f_{kw} = R_s$, and for each relay $i$,
\[
\sum_{h \in {\cal P}_i} f_{hi} = \sum_{j \in {\cal O}_i} f_{ij} \triangleq r_i,
\]
where $r_i$ denotes the incoming flow rate at $i$.

\subsection{Link Cost and Pricing Functions}\label{sec:CostFunction}

Each link has a strictly increasing and strictly convex cost
function $D_{ij}(f_{ij})$, which is \emph{private information} to
$i$ and $j$ only. For example, $D_{ij}(f_{ij})$ can represent the
queuing delay incurred on $(i,j)$ with arrival rate $f_{ij}$, e.g.
the average occupancy function $f_{ij}/(c_{ij} - f_{ij})$ of an
M/M/1 queue with service rate $c_{ij}$. As another example, if the
links are wireless, $D_{ij}(f_{ij})$ can measure the transmission
power required for achieving rate $f_{ij}$.  For example, if the
link transmission rate $f_{ij}$ is determined by transmission power
$P_{ij}$ as $f_{ij} = W \log(1 + K P_{ij})$ for some constants $W, K
> 0$,\footnote{Assume that with proper time or frequency scheduling,
transmission on different links are non-interfering.} then $ P_{ij}
= \frac{1}{K} \left(2^{f_{ij}/W} - 1 \right) \triangleq
D_{ij}(f_{ij})$, which is strictly increasing and convex in
$f_{ij}$. As suggested by the examples, the analytical framework
presented above applies to both wireline and wireless networks.

For analytical purposes, we further assume that $D_{ij}(\cdot)$ is continuously
differentiable with derivative $d_{ij}(\cdot)$. By previous assumptions, $d_{ij}(\cdot)$
is positive and strictly increasing.  The \emph{socially optimal} routing is the routing
that minimizes the network cost $\sum_{(i,j)}D_{ij}(f_{ij})$. Because link costs are
strictly convex, the socially optimal routing is uniquely characterized by the condition
that every path from $s$ to $w$ with positive flow\footnote{The flow rate of a path is
the minimum of the flow rates of all the links on that path.} has the minimum marginal
cost among all paths. For otherwise, one can reduce the total cost by shifting an
infinitesimal amount of flow from a path with non-minimum marginal cost to a
minimum-marginal-cost path.

We model the source and relays as selfish agents who must pay for
the costs on their outgoing links. While the source has to send all
its traffic out, it strives to do this with the minimum cost. On the
other hand, a relay has an incentive to forward traffic for its
predecessors only if it is adequately rewarded for its service in
the form of payment by its predecessors. The amount of payment is
determined as follows.

Suppose a node $h$ has incoming flow of rate $r_h > 0$. Each $i \in
{\cal O}_h$ announces a pricing function $P_i^h(\cdot)$ which
specifies the payment $P_i^h(f_{hi})$ it demands should $h$ forward
traffic of rate $f_{hi}$ to it.\footnote{The domain of
$P_i^h(\cdot)$ must contain the interval $[0, r_h]$. And as we will
see, a selfish and strategic relay $i$ always designs $P_i^h(\cdot)$
tailored to the total traffic $r_h$. However, for simplicity we do
not express such dependence in the notation.} For analytical
purposes, we assume that $P_i^h(t)$ is continuously differentiable
with derivative $p_i^h(t)$. Note that $P_i^h(\cdot)$ provides $h$ a
{\em continuum} of options, namely the rate-price pairs $(f_{hi},
P_i^h(f_{hi}))$.\footnote{If $i$ has multiple predecessors,
$P_i^h(t)$ for one $h \in {\cal P}_i$ is an agreement exclusively
between $i$ and $h$, independent of the flow rates allocated to $i$
by other predecessors. Presumably, however, $i$ designs $P_i^h(t)$
for all $h \in {\cal P}_i$ jointly because $r_i=\sum_{h \in {\cal
P}_i}f_{hi}$, and $i$ has to pay its offsprings to get $r_i$
forwarded.} After learning $(P_i^h(\cdot))_{i \in {\cal O}_h}$, $h$
decides on the allocation of $r_h$ and makes payments to its
offsprings accordingly.

\subsection{Pricing Game}

We assume every node is selfish and strategic.\footnote{The
destination $w$ is the only node that plays no active role in the
pricing game described below. It passively accepts the flow assigned
by its predecessors, who treat it as an offspring using a
uniformly-zero pricing function. Because $\cal G$ is assumed to be
loop-free and provide directed path(s) to $w$ from every other node,
the total flow arriving at $w$ must be equal to $R_s$.} The source
$s$ thus always allocates the total flow $R_s$ to the nodes in
${\cal O}_s$ so as to minimize its total cost, which includes the
costs on its outgoing links and the payments to its offsprings.
Specifically, given the pricing functions $P_i^s(\cdot)$ of $i \in
{\cal O}_s$, the optimal allocation of $R_s$ from the perspective of
$s$ is any \be\label{eq:SourceOptRouting} (f_{si}^*)_{i \in {\cal
O}_s} \in \mathop{\arg\min}_{(f_{si}) \in {\cal F}_s(R_s)} \sum_{i
\in {\cal O}_s} D_{si}(f_{si}) + P_i^s(f_{si}), \ee where ${\cal
F}_i(r)$ is defined as $\{(f_{ik})_{k \in {\cal O}_i} \ge \bs 0:
\sum_k f_{ik} = r\}$.

A relay $i$ is a predecessor to some nodes, and is an offspring to some other nodes. As a predecessor, it acts just
like $s$. That is, it allocates the total incoming flow in the most cost efficient way from its own perspective. Thus,
the traffic allocation adopted by $i$ when it has incoming flow $r_i$ is any \be\label{eq:RelayOptRouting}
(f_{ik}^*)_{k \in {\cal O}_i} \in \mathop{\arg\min}_{(f_{ik}) \in {\cal F}_i(r_i)} \sum_{k \in {\cal O}_i}
D_{ik}(f_{ik}) + P_k^i(f_{ik}). \ee Denote the minimum value in~\eqref{eq:RelayOptRouting} by $D_i(r_i)$. Note that
$D_i(r_i)$ represents the minimum cost to $i$ for forwarding flow of rate $r_i$.\footnote{Although not explicit from
the notation, $D_i(\cdot)$ depends on $(P_k^i(\cdot))_{k \in {\cal O}_i}$.} It is easy to show that $D_i(\cdot)$ is
continuous and increasing with piecewise continuous derivative denoted by $d_i(\cdot)$.

As an offspring, $i$ designs $P_i^h(\cdot)$ for every $h \in {\cal
P}_i$ with the aim of maximizing its profit in competition with its
siblings (discussed in depth later). It does this with the
assumption that $h$ always allocates in the most cost efficient way,
and that $r_h$ for each $h \in {\cal P}_i$ stays constant at the
current value irrespective of its choice of $P_i^h(\cdot)$. While
the first assumption is very reasonable, the second one requires
some justification.

Theoretically, since $r_h$ of $h \in {\cal P}_i$ is the outcome of the optimal allocation by $h$'s predecessors, it in
general cannot stay constant if $h$ changes its pricing functions. However, $h$'s pricing functions presumably are tied
to $i$'s choice of $P_i^h(\cdot)$ as the total cost to $h$ is partly leveraged by the price charged by $i$. Once $h$
reacts to the change in $P_i^h(\cdot)$ by updating its own pricing functions, $r_h$ is inevitably adjusted by $h$'s
predecessors. On the other hand, relays are usually too myopic to note this chain reaction since they have very limited
knowledge of the network. Recall that we made a practical assumption that a relay is aware of only its predecessors,
siblings and offsprings. As a result, it can at best predict the impact of its strategy on the traffic allocation by
its predecessors, but not nodes further upstream. It is therefore reasonable for $i$ to consider only the competition
with its siblings for the flow their common predecessors currently have.

We now formally define the (static) pricing game (PG) as having the following components:

\bi

\item The set of players ${\cal I} = {\cal N}\backslash\{s,w\}$: relays in ${\cal G}$.

\item Strategy of player $i$: continuously differentiable pricing functions $P_i^h(\cdot)$ for all $h \in {\cal P}_i$.

\item Payoff to player $i$: the profit made by servicing $(f_{hi}^*)_{h \in {\cal P}_i}$:
\be\label{eq:PGPayoff} \sum_{h \in {\cal P}_i} P_i^h(f_{hi}^*) - D_i\left(\sum_{h \in {\cal
P}_i}f_{hi}^*\right), \ee where the routing $(f_{ij}^*)_{(i,j)\in{\cal E}}$ is most cost efficient
from the perspective of every node, i.e., \eqref{eq:SourceOptRouting}-\eqref{eq:RelayOptRouting}
hold for $s$ and every relay $i$ where $r_i = \sum_{h \in {\cal P}_i} f_{hi}^*$.

\ei

A pricing game is fully characterized by the tuple $({\cal G},
(D_{ij}(\cdot)), R_s)$. In the rest of the paper, we will study the
outcome of the PG with \emph{myopic} players as described above. The
focus of our work is to investigate whether the PG has an
equilibrium where no relay can increase its profit by unilaterally
changing its pricing functions, and when an equilibrium exists, how
the resulting routing compares to the socially optimal one.

\subsection{Best Response and Equilibrium}\label{subsec:Equilibrium}

Each player $i$ in the PG is assumed to be myopic in the sense that
it knows the pricing functions of all its competitors as well as its
downstream nodes. Based on its local information $\bs L_i \triangleq
((r_h, (P_j^h(\cdot))_{j \in {\cal S}_i^h})_{h \in {\cal P}_i},
(P_k^i(\cdot))_{k \in {\cal O}_i})$, player $i$ anticipates a payoff
$\Gamma_i(\bs P_i; \bs L_i)$ when adopting $\bs P_i \triangleq
(P_i^h(\cdot))_{h \in {\cal P}_i}$, where
\[
\Gamma_i(\bs P_i; \bs L_i) \triangleq \sum_{h \in {\cal P}_i} P_i^h(f_{hi}^*) - \sum_{k
\in {\cal O}_i} \left[ D_{ik}(f_{ik}^*) + P_k^i(f_{ik}^*) \right],
\]
$(f_{hj}^*)_{j \in {\cal O}_h}$ is the optimal allocation of $r_h$
by $h \in {\cal P}_i$ given $(P_{j}^h(\cdot))_{j \in {\cal O}_h}$,
and $(f_{ik}^*)_{k \in {\cal O}_i}$ is the optimal allocation of
$r_i^* = \sum_{h \in {\cal P}_i} f_{hi}^*$ given
$(P_k^i(\cdot))_{k \in {\cal O}_i}$. 

\vspace{0.1in}\begin{definition}\label{def:BestResp} A pricing function profile
$(P_i^h(\cdot))_{h \in {\cal P}_i}$ is a best response to local information $\bs L_i$ if
\[
\Gamma_i(\bs P_i; \bs L_i) = \max_{\bs Q_i \textrm{feasible}} \Gamma_i(\bs Q_i; \bs L_i),
\]
where $\bs Q_i = (Q_i^h(\cdot))_{h \in {\cal P}_i}$ is feasible if every component is a
continuously differentiable function.
\end{definition}\vspace{0.1in}

Denote the set of best responses to $\bs L_i$ by ${\cal B}_i(\bs
L_i)$. Before we proceed, we must prove that ${\cal B}_i(\bs L_i)$
is non-empty.

\vspace{0.1in}\begin{lemma}\label{lma:BestResp}  For any $\bs L_i$,
the set ${\cal B}_i(\bs L_i)$ is non-empty. Furthermore, $\bs P_i
\in {\cal B}_i(\bs L_i)$ if and only if for all $h \in {\cal P}_i$
and all $t \in [0, r_h]$, \be\label{eq:BestResponse1} B_i^h(t)
\triangleq D_{hi}(t) + P_i^h(t) \ge B_{\hat i}^h(r_h) - B_{\hat
i}^h(r_h - t), \ee and \be\label{eq:BestResponse2} B_i^h(\tilde
f_{hi}) = B_{\hat i}^h(r_h) - B_{\hat i}^h(r_h - \tilde f_{hi}), \ee
where \be\label{eq:VirtualComp} B_{\hat i}^h(r) \triangleq
\min_{\substack{(f_{hj}) \in {\cal F}_h(r)\\f_{hi}=0}} \sum_{j \in
{\cal S}_i^h} B_j^h(f_{hj}) \ee and $(\tilde f_{hi})_{h \in {\cal
P}_i}$ is a vector that maximizes
\[
\bar\Gamma_i(\bs f_i; \bs L_i) \triangleq \sum_{h \in {\cal P}_i} \left[B_{\hat i}^h(r_h) - B_{\hat i}^h(r_h - f_{hi})
- D_{hi}(f_{hi}) \right] - D_i\left(\sum_{h \in {\cal P}_i}f_{hi}\right)
\]
over all the $\bs f_i = (f_{hi})$ such that $0 \le f_{hi} \le r_h$
for all $h \in {\cal P}_i$.
\end{lemma}\vspace{0.1in}

Before giving the proof, we first provide some intuitive
explanations for the lemma. The function $B_i^h(t)$ gives the total
cost $h$ spends on routing traffic of rate $t$ to $i$. Since
$D_{hi}(\cdot)$ is fixed and known to $i$, it is equivalent to treat
$B_i^h(\cdot)$ as the pricing function $i$ uses to charge $h$. With
this view, $h$ makes a lump-sum payment to $i$ determined by
$B_i^h(\cdot)$, and lets $i$ pay for the cost $D_{hi}(\cdot)$ on
link $(h,i)$. For convenience, from now on we assume each relay $i$
announces $B_i^h(\cdot)$ to $h \in {\cal P}_i$ and siblings $j \in
{\cal S}_i^h$. By~\eqref{eq:VirtualComp}, $B_{\hat i}^h(r)$
represents the minimum cost $h$ can achieve by forwarding traffic of
rate $r$ to offsprings other than $i$. It will become evident in the
next proof that from $i$'s viewpoint, the competition from all $j\in
{\cal S}_i^h$ can be aggregated into a virtual competitor $\hat i^h$
using pricing function $B_{\hat i}^h(\cdot)$. Thus, it is as if $i$
were competing with one relay $\hat i^h$ in each ``market'' $h \in
{\cal P}_i$. The vector $(\tilde f_{hi})_{h \in {\cal P}_i}$
represents the ``market shares'' that jointly yield the maximum
(anticipated) profit to $i$. Pricing functions $B_i^h(\cdot)$ which
satisfy the conditions in Lemma~\ref{lma:BestResp} induce $h \in
{\cal P}_i$ to allocate the ideal ``market share'' $\tilde f_{hi}$
to $i$ and give $i$ the maximum profit. This is because
\eqref{eq:BestResponse2} implies that allocating $\tilde f_{hi}$ to
$i$ and the rest to other relays yields the same cost to $h$ as
allocating all the traffic to other relays. So
conditions~\eqref{eq:BestResponse1} and \eqref{eq:BestResponse2}
combined imply that no other allocation costs less than the above
two schemes.


{\vspace{0.1in}\textit{Proof of Lemma~\ref{lma:BestResp}:} First notice that for fixed $\bs L_i$,
the profit of $i$ when it adopts $(B_i^h(\cdot))_{h \in {\cal P}_i}$ is upper bounded as follows:
\beas \Gamma_i((B_i^h(\cdot)); \bs L_i) &=& \sum_{h \in {\cal P}_i} \left[
B_i^h(f_{hi}) - D_{hi}(f_{hi})\right] - D_i\left(\sum_{h \in {\cal P}_i}f_{hi}\right) \\
&\le& \sum_{h \in {\cal P}_i} \left[B_{\hat i}^h(r_h) - B_{\hat i}^h(r_h - f_{hi})-
D_{hi}(f_{hi})\right] - D_i\left(\sum_{h \in {\cal P}_i}f_{hi}\right) \\
&=& \bar\Gamma_i(\bs f_i; \bs L_i) \le \bar\Gamma_i(\bs {\tilde
f_i}; \bs L_i), \eeas where $\bs f_i \triangleq (f_{hi})_{h \in
{\cal P}_i}$ is the optimal amount of traffic allocated to $i$ when
$i$ uses $(B_i^h(\cdot))$. The first inequality holds because for
every $h \in {\cal P}_i$, $B_i^h(f_{hi}) \le B_{\hat i}^h(r_h) -
B_{\hat i}^h(r_h - f_{hi})$. For otherwise, $h$ would find the cost
of allocating $r_h$ exclusively to all $j \in {\cal S}_i^h$
($B_{\hat i}^h(r_h)$) strictly less than the cost of allocating
$f_{hi}$ to $i$ and $r_h - f_{hi}$ to $j \in {\cal S}_i^h$
($B_i^h(f_{hi}) + B_{\hat i}^h(r_h - f_{hi})$), a contradiction. The
second inequality follows from the definition of $\tilde{{\bs
f}_i}$.}

{Notice that the upper bound $\bar\Gamma_i(\bs {\tilde f_i}; \bs L_i)$ is independent of
$(B_i^h(\cdot))$. It is tight if and only if both inequalities hold with equality. To make the
second inequality tight, it is necessary and sufficient to have the traffic allocation $\bs f_i$
induced by $(B_i^h(\cdot))$ equal to $\bs {\tilde f_i}$. Given that, the first inequality is tight
if \eqref{eq:BestResponse1}-\eqref{eq:BestResponse2} hold, since they guarantee that allocating
$\tilde f_{hi}$ to $i$ and $r_h - \tilde f_{hi}$ to $j \in {\cal S}_i^h$ is in $h$'s best interest.
They are also necessary because \eqref{eq:BestResponse2} is prerequisite for the first inequality
to be tight, and consequently \eqref{eq:BestResponse1} cannot be violated either. For example, if
for some $h \in {\cal P}_i$ and some $t \in [0, r_h]$, $B_i^h(t) < B_{\hat i}^h(r_h) - B_{\hat
i}^h(r_h - t)$. Then allocating $t$ to $i$ and $r_h - t$ to $j \in {\cal S}_i^h$ would incur
strictly less cost to $h$ than $B_{\hat i}^h(r_h) = B_i^h(t)(\tilde f_{hi}) + B_{\hat i}^h(r_h -
\tilde f_{hi})$. Thus, $h$ would not have allocated $\tilde f_{hi}$ to $i$, which is a
contradiction. \qed}

Note that best response $(B_i^h(\cdot))$ always exists because, for instance, $B_i^h(t) = B_{\hat
i}^h(r_h) - B_{\hat i}^h(r_h - t)$ for $t \in [0, r_h]$ satisfies
\eqref{eq:BestResponse1}-\eqref{eq:BestResponse2}. The (pure-strategy) Nash equilibria are defined
as the fixed points of the best response mapping.

\vspace{0.1in}\begin{definition}\label{def:NE} Pricing function
profiles $\bs P_i, i \in {\cal I}$ constitute an equilibrium if for
all $i \in {\cal I}$, $\bs P_i \in {\cal B}_i(\bs L_i)$ where the
incoming flow vector $(r_h)_{h \in {\cal P}_i}$ contained in $\bs
L_i$ results from the routing $(f_{ij}^*)_{(i,j)\in{\cal E}}$ that
is most cost efficient from the perspective of every individual
node, i.e., \eqref{eq:SourceOptRouting}-\eqref{eq:RelayOptRouting}
hold for $s$ and all $i \in {\cal I}$,
\[ r_h = \sum_{g \in {\cal P}_h} f_{gh}^*, \quad\textrm{if}~ h \ne s
\]
and $r_h = R_s$ if $h = s$.
\end{definition}\vspace{0.1in}

It is easy to see that if $(\bs P_i)_{i \in {\cal I}}$ constitutes
an equilibrium, $\Gamma_i(\bs P_i; \bs L_i)$ must coincide with the
actual payoff of $i$.

\vspace{0.1in}\begin{definition}\label{def:EfficientNE} An
equilibrium $(\bs P_i)_{i \in {\cal I}}$ is \emph{efficient} if it induces
the socially optimal routing. In this case, $(\bs P_i)_{i \in {\cal
I}}$ is said to induce the social optimum.
\end{definition}\vspace{0.1in}

Before proving the existence of equilibria and analyzing their efficiency in the general setting, we first study pricing games under some simple network
topologies. The analysis of these games not only provide valuable insight into the general problem
but also have significant implications in their own right.

\section{Equilibria in Oligopoly}\label{sec:Oligopoly}

\begin{figure}[!h]
  \begin{center}
  \includegraphics[width=5cm]{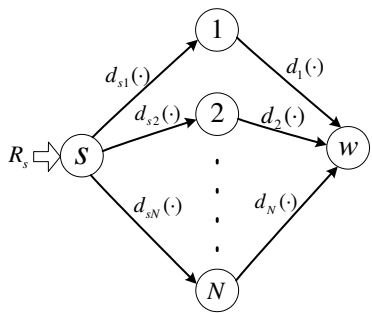}\\
  \caption{Oligopoly with N relays.}\label{fig:Oligopoly}
  \end{center}
\end{figure}

The simplest topologies within our framework are those including a single layer of relays, e.g. the
one in Figure~\ref{fig:Oligopoly}.
Here, $N$ relays each have a direct link from $s$ and a direct link
to $w$. They compete for the total flow $R_s$ by advertising their
pricing functions $\beta_i(\cdot) = d_{si}(\cdot) + p_i(\cdot), i =
1, \cdots, N$. From now on, we will more often refer to the
derivatives $d_{ij}(\cdot)$ and $p_i(\cdot)$ as link cost and
pricing functions, since they appear to be more convenient for
marginal cost analysis at equilibria. Also we will use simplified
notation whenever appropriate, e.g. the superscript is omitted from
$p_i(\cdot), \beta_i(\cdot)$ as $s$ is the only predecessor to every
relay.  We refer to a pricing game under such a topology as an
\emph{oligopoly}. Define $\lambda_i(t)\triangleq d_{si}(t) +
d_i(t)$. An oligopoly PG is fully characterized by the tuple $(N,
(\lambda_i(\cdot))_{i=1}^N, R_s)$.

Because all $\lambda_i(\cdot)$ are strictly increasing, the socially
optimal routing $(r_i^*)_{i = 1}^N$ is unique and is given by
\[
\lambda_i(r_i^*) = \min_{j = 1, \cdots, N} \lambda_j(r_j^*)
\]
if $r_i^* > 0$.

We now analyze the routing established by the oligopoly PG. Given
$(\beta_i(\cdot))_{i \in {\cal O}_s}$, the self-interest of $s$
leads it to adopt the most cost efficient routing
$(f_{si}^*)_{i=1}^N$ such that \be\label{eq:OligSourceOpt}
\beta_i(f_{si}^*) = \min_{j = 1, \cdots, N} \beta_j(f_{sj}^*) \ee
whenever $f_{si}^* > 0$. Whether $(f_{si}^*) = (r_i^*)$ or not
depends on how $(\beta_i(\cdot))$ are chosen by the individual
relays.

\subsection{Best Response and Existence of Equilibria}\label{subsec:OligBestResponse}

\begin{figure}[!h]
  \begin{center}
  \includegraphics[width=5cm]{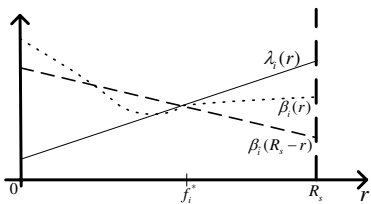}\\
  \caption{Typical best response curve in oligopoly.}\label{fig:OligBestResponse}
  \end{center}
\end{figure}

We apply Lemma~\ref{lma:BestResp} to the oligopoly PG. Here,
\[
B_i(t) = \int_0^t \beta_i(r)~dr,
\]
\[
B_{\hat i}(t) = \min_{\sum_{j \ne i} f_{sj} = t} \sum_{j \ne i}
\int_0^{f_{sj}} \beta_j(r)~dr.
\]
It is easy to show that $B_{\hat i}(t)$ is continuous and increasing. Its derivative, denoted by $\beta_{\hat i}(t)$,
is in general piecewise continuous. For $t \in (0, R_s)$, let the left and right limits of $\beta_{\hat i}(t)$ be
denoted by $\beta_{\hat i}(t)^-$ and $\beta_{\hat i}(t)^+$.\footnote{It is understood that $\beta_{\hat i}(0)$ has only
a right limit and that $\beta_{\hat i}(R_s)$ has only a left limit.} By Lemma~\ref{lma:BestResp}, the best response of
$i$ given $\beta_{\hat i}(\cdot)$ can be simply characterized by \be\label{eq:OligBestResponse1} \int_0^{t}
\beta_i(r)~dr \left\{\ba{ll}
                                \ge \int_0^{t} \beta_{\hat i}(R_s -
                                r)~dr, \quad &0\le t \le R_s\\
                                = \int_0^{t} \beta_{\hat i}(R_s -
                                r)~dr, &t = f_i^*, \ea \right.
\ee where \be\label{eq:OligBestResponse2} f_i^* \in \mathop{\arg\max}_{0 \le f_i \le R_s} \int_0^{f_i} \beta_{\hat
i}(R_s - r) - \lambda_i(r)~dr. \ee To gain an intuitive idea of the above conditions, suppose $\beta_{\hat i}(R_s - r)$
and $\lambda_i(r)$ are given by the dashed and solid curves in Figure~\ref{fig:OligBestResponse}. A typical best
response $\beta_i(r)$ is shown as the dotted curve.
In particular, one can let $\beta_i(r)$ coincide with $\beta_{\hat i}(R_s - r)$ on $[0, f_i^*]$ and let $\beta_i(r) \ge
\beta_{\hat i}(R_s - r)$ on $(f_i^*, R_s]$. Such a best response will be referred to as a \emph{replicating} response.
As we will show, oligopoly equilibria induced by replicating responses are always efficient while equilibria induced by
other best responses are not necessarily efficient.

\subsection{Efficient Equilibria}

\begin{theorem}\label{thm:OligOpt}

The socially optimal routing of an oligopoly can always be induced by an equilibrium.

\end{theorem}\vspace{0.1in}

\textit{Proof:} We prove the theorem by constructing an equilibrium
that induces the socially optimal routing $(r_i^*)$. Define
$\lambda^* \triangleq \min_{j = 1, \cdots, N} \lambda_j(r_j^*)$. Let
$\beta_i(r) \equiv \lambda^*$ for all $i$. Then $\beta_i(r) =
\beta_{\hat i}(R_s - r) = \lambda^*$ is a best response for all $i$
with $f_i^* = r_i^*$. Thus, $(\beta_i(\cdot))$ constitutes an
equilibrium which results in the routing $(r_i^*)$. \qed

Because the socially optimal routing always exists, we can conclude that there \emph{always} exists an \emph{efficient}
equilibrium for any oligopoly pricing game.

\begin{figure}[!h]
  \begin{center}
  \includegraphics[width=5cm]{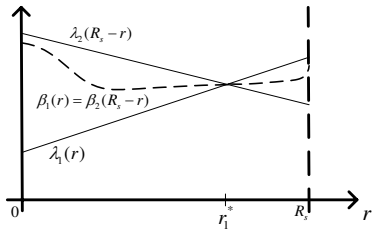}\\
  \caption{General (focal) equilibrium in duopoly.}\label{fig:DuopolyFocalEqui}
  \end{center}
\end{figure}

Although we used constant $(\beta_i(\cdot))$ (or linear pricing functions $(B_i(\cdot))$) to construct an efficient
equilibrium in the proof, efficient equilibria can be established by nonlinear pricing functions as well. For instance,
Figure~\ref{fig:DuopolyFocalEqui} depicts an equilibrium in a duopoly PG where the two relays adopt $\beta_1(\cdot),
\beta_2(\cdot)$ of a more general shape. Notice that in a duopoly, $\beta_{\hat 1}(t) = \beta_2(t)$ and $\beta_{\hat
2}(t) = \beta_1(t)$.

To derive a simple criterion for checking the efficiency of an equilibrium, we need to make the following distinction.
A routing $(f_i)_{i=1}^N$ is said to be \emph{monopolistic} if $f_m = R_s$ for some relay $m$ and $f_j = 0$ for all $j
\ne m$. In this case, $m$ is called the \emph{dominant} relay. An equilibrium is \emph{monopolistic} if it induces a
monopolistic routing. A routing is said to be \emph{competitive} if there are at least two relays $i,j$ such that
$f_i>0, f_j>0$. An equilibrium is \emph{competitive} if it induces a competitive routing.
\vspace{0.1in}\begin{theorem}\label{thm:OligCompEqui} If an oligopoly equilibrium is competitive, it must be efficient.
\end{theorem}\vspace{0.1in}

We will need the next lemma to prove Theorem~\ref{thm:OligCompEqui}.

\vspace{0.1in}\begin{lemma}\label{lma:OligVirtualMargin}
At an
oligopoly equilibrium $(\beta_i(\cdot))$ which induces routing
$(f_i^*)$, if $0 < f_i^* \le R_s$, then for all $j \ne i$,
\[ \beta_{\hat i}(R_s - f_i^*)^+ \le \beta_j(f_j^*).
\]
If $0 \le f_i^* < R_s$, then for all $j \ne i$ such that $f_j^*>0$,
\[ \beta_{\hat
i}(R_s - f_i^*)^- \ge \beta_j(f_j^*).
\]
\end{lemma}\vspace{0.1in}

{\textit{Proof:} By definition, if $t<R_s$, $\beta_{\hat i}(t)^+ = \lim_{\Delta\to 0^+} (B_{\hat
i}(t+\Delta)-B_{\hat i}(t))/\Delta$. Therefore when $f_i^*>0$, \beas\beta_{\hat i}(R_s -
f_i^*)^+&=& \lim_{\Delta\to 0^+} \frac{1}{\Delta}\left\{B_{\hat i}(R_s - f_i^*+\Delta) -
B_{\hat i}(R_s - f_i^*)\right\}\\
&\stackrel{(a)}{=}& \lim_{\Delta\to 0^+}\frac{1}{\Delta}\left\{\min_{\sum_{j\ne i}f_j = R_s-f_i^*+\Delta}\sum_{j\ne
i}\int_0^{f_j} \beta_j(r)~dr-\sum_{j\ne
i}\int_0^{f_j^*} \beta_j(r)~dr\right\}\\
&\stackrel{(b)}{\le}& \lim_{\Delta\to 0^+}\frac{1}{\Delta}\left\{ \int_0^{f_k^*+\Delta} \beta_k(r)~dr + \sum_{j\ne
i,k}\int_0^{f_j^*} \beta_j(r)~dr -
\sum_{j\ne i}\int_0^{f_j^*} \beta_j(r)~dr\right\}\\
&=& \lim_{\Delta\to 0^+}\frac{1}{\Delta} \int_{f_k^*}^{f_k^*+\Delta} \beta_k(r)~dr = \beta_k(f_k^*).\eeas Here,
equation (a) follows from the fact that $(f_j^*)_{j=1}^N$ is the equilibrium routing of $R_s$ induced by $
(\beta_j(\cdot))_{j=1}^N$. Inequality (b) is obtained by substituting the minimum-cost routing of $R_s-f_i^*+\Delta$ by
an arbitrary routing, namely $f_k^*+\Delta$ allocated to $k$ and $f_j^*$ to each $j\ne i,k$. The second inequality in
the lemma can be proved in a similar manner.\qed }

\textit{Proof of Theorem~\ref{thm:OligCompEqui}:} Let $(f_i^*)$ be the routing induced by a competitive equilibrium
$(\beta_i(\cdot))$. Let $m,n$ be any two relays such that $f_m^*>0, f_n^*>0$. It is enough to show that
$\lambda_m(f_m^*) = \lambda_n(f_n^*)$ and that $\lambda_m(f_m^*) \le \lambda_j(f_j^*)$ for any $j$ with $f_j^*=0$.
By~\eqref{eq:OligSourceOpt}, $\beta_m(f_m^*) = \beta_n(f_n^*)$. The best response condition~\eqref{eq:OligBestResponse2}
implies that $\beta_{\hat m}(R_s - f_m^*)^- \le \lambda_{m}(f_m^*) \le \beta_{\hat m}(R_s - f_m^*)^+$. By
Lemma~\ref{lma:OligVirtualMargin}, $\beta_n(f_n^*)\le\beta_{\hat m}(R_s - f_m^*)^-$ and $\beta_{\hat m}(R_s - f_m^*)^+
\le \beta_n(f_n^*)$. In conclusion, $\beta_n(f_n^*) = \lambda_{m}(f_m^*)$. By symmetry, we can show that
$\beta_m(f_m^*) = \lambda_{n}(f_n^*)$. Therefore, $\lambda_m(f_m^*) = \lambda_n(f_n^*)$. Now suppose $f_j^*=0$.
By~\eqref{eq:OligBestResponse2} and Lemma~\ref{lma:OligVirtualMargin}, $\lambda_j(0) \ge \beta_{\hat j}(R_s)^- \ge
\beta_n(f_n^*) = \lambda_m(f_m^*)$. So the proof is complete. \qed

\subsection{Inefficient Equilibria}\label{subsec:IneffiEqui}

\begin{figure}[!h]
  \begin{center}
  \includegraphics[width=5cm]{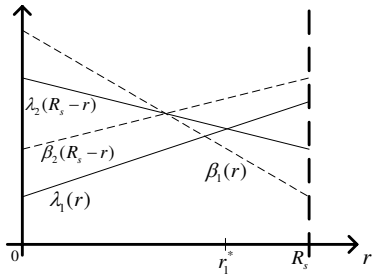}\\
  \caption{Inefficient equilibrium in duopoly.}\label{fig:BadDuopoly}
  \end{center}
\end{figure}

Theorem~\ref{thm:OligCompEqui} does not rule out the possibility of inefficient equilibria. In fact,
an equilibrium may be inefficient if it is monopolistic. For example, the socially optimal routing
of the duopoly PG represented by Figure~\ref{fig:BadDuopoly} is $(r_1^*, R_s - r_1^*)$ whereas the
equilibrium depicted leads to a monopolistic routing $(R_s,0)$.
In this example, relay $2$ adopts a pricing function $\beta_2(\cdot)$ such that $\int_0^{R_s} \beta_2(r)~dr =
\int_0^{R_s} \lambda_2(r)~dr$ and $\beta_2(R_s - r) > \lambda_1(r)$ for all $r \in [0,R_s]$. Given such a
$\beta_2(\cdot)$, relay $1$ would want to acquire all the flow (cf.~\eqref{eq:OligBestResponse2}) by using
$\beta_1(\cdot)$ such that $\int_0^t \beta_1(R_s - r)~dr < \int_0^t \lambda_2(r)~dr$ and $\int_0^t \beta_1(r)~dr >
\int_0^t \beta_2(R_s-r)~dr$ for all $t \in (0,R_s)$ and $\int_0^{R_s} \beta_1(R_s - r)~dr = \int_0^{R_s}
\lambda_2(r)~dr = \int_0^{R_s} \beta_2(r)~dr$. Thus, it satisfies~\eqref{eq:OligBestResponse1} and leaves relay $2$ no
incentive to acquire any traffic. So the monopolistic equilibrium holds.

In general, monopolistic equilibria in an oligopoly PG have the following property.
\vspace{0.1in}\begin{theorem}\label{thm:OligMonoEqui} If an oligopoly equilibrium is monopolistic with dominant
relay $m$, we must have
\[
\int_0^{R_s} \lambda_m(r)~dr \le \int_0^{R_s} \lambda_j(r)~dr
\]
for any other relay $j$.
\end{theorem}\vspace{0.1in}

{\textit{Proof:} Consider any $j$ with $f_j^* = 0$ in a monopolistic equilibrium. The
condition~\eqref{eq:OligBestResponse2} implies that
\[
\int_0^{R_s} \lambda_j(r)~dr \ge \int_0^{R_s} \beta_{\hat j}(R_s -
r)~dr.
\]
On the other hand,
\[
\int_0^{R_s} \beta_{\hat j}(R_s - r)~dr = \int_0^{R_s} \beta_{\hat j}(r)~dr = \int_0^{R_s} \beta_m(r)~dr,
\]
since from the perspective of $s$, the optimal allocation of $R_s$ to all the relays except $j$ still assigns all the
traffic to $m$. It follows from $m$'s best response
conditions~\eqref{eq:OligBestResponse1}-\eqref{eq:OligBestResponse2} that
\[
\int_0^{R_s} \beta_m(r)~dr = \int_0^{R_s} \beta_{\hat m}(R_s - r)~dr \ge \int_0^{R_s} \lambda_m(r)~dr.
\]
Thus the proof is complete. \qed}

The next conclusion easily follows from Theorem~\ref{thm:OligMonoEqui}.
\vspace{0.1in}\begin{corollary}\label{cor:OligOpt} If the socially optimal routing of an oligopoly is monopolistic,
then every equilibrium of the oligopoly is monopolistic and efficient.
\end{corollary}\vspace{0.1in}

{\textit{Proof:} It can deduced from the uniqueness of the socially optimal routing and
Theorem~\ref{thm:OligCompEqui} that every equilibrium of such an oligopoly must be monopolistic. By
Theorem~\ref{thm:OligMonoEqui}, the dominant relay $m$ of such an equilibrium has the minimum
$\int_0^{R_s} \lambda_m(r)~dr$ among all relays. But such an $m$ must be the dominant relay in the
socially optimal routing. \qed}

It is shown next that there always exists a monopolistic equilibrium
in an oligopoly. Thus, we have the following conclusion.
\begin{corollary}\label{cor:OligInefficient}
If the socially optimal routing of an oligopoly is competitive, then there exists an inefficient (monopolistic)
equilibrium.
\end{corollary}

{\textit{Proof:} We need only show that there exists a monopolistic equilibrium in such an
oligopoly. Let all $\beta_j(\cdot)$ be the same strictly decreasing function $\beta(\cdot)$ such
that $\int_0^t \beta(R_s - r) dr \le \int_0^t \lambda_j(r) dr$ for all $j$ and $t \in [0,R_s)$ but
$\int_0^{R_s} \beta(R_s - r) dr = \int_0^{R_s} \lambda_m(r) dr$ where $m \in \arg\min_{j}
\int_0^{R_s} \lambda_j(r) dr$. Since $\beta(\cdot)$ is strictly decreasing, $\beta_{\hat j}(r) =
\beta(r)$ for all $j$. By construction, $f_j^* = 0$ is an ideal flow to $j \ne m$
(cf.~\eqref{eq:OligBestResponse2}) whereas $\beta_m(\cdot) = \beta(\cdot)$ and $f_m^* = R_s$
jointly satisfy $m$'s best response
conditions~\eqref{eq:OligBestResponse1}-\eqref{eq:OligBestResponse2}. So the monopolistic
equilibrium is established. \qed}


When an oligopoly has inefficient equilibria, it is of interest to compare the worst-case network cost under an
inefficient equilibrium to the optimal cost.

\subsection{Price of Anarchy}

The price of anarchy, as a measure of loss of social efficiency due
to selfish behavior of individual agents, was studied in the
literature on selfish routing~\cite{paper:Rou02,paper:Rou05}. In
this work, the price of anarchy of a general PG is defined as
follows. \vspace{0.1in}\begin{definition}\label{def:POA} The price
of anarchy $\rho({\cal G}, (D_{ij}(\cdot)), R_s)$ of a pricing game
$({\cal G}, (D_{ij}(\cdot)), R_s)$ is the ratio of the maximum cost
at an equilibrium to the socially optimal cost, i.e.,
\[
\rho({\cal G}, (D_{ij}(\cdot)), R_s) \triangleq \frac{\max_{(f_{ij}) \in {\cal
F}^E}\sum_{(i,j)\in{\cal E}}D_{ij}(f_{ij})}{\sum_{(i,j)\in{\cal E}}D_{ij}(f_{ij}^*)},
\]
where ${\cal F}^E$ is the collection of all routings that can be induced by an equilibrium of $({\cal G},
(D_{ij}(\cdot)), R_s)$ and $(f_{ij}^*)_{(i,j)\in{\cal E}}$ is the socially optimal routing of the game.
\end{definition}\vspace{0.1in}

In this section, we study the price of anarchy specifically for
oligopolies. As we will show, $\rho(N, (\lambda_i(\cdot)), R_s)$ is
equal to $N$ when marginal cost functions are concave, e.g. when
cost functions are quadratic. However, the price of anarchy can be
arbitrarily large when when marginal cost functions are convex, as
is the case for the cost functions discussed in
Section~\ref{sec:CostFunction}.

\vspace{0.1in}\begin{theorem}\label{thm:ConcavePOA} If the cost derivatives $(\lambda_i(\cdot))$ are concave, $\rho(N,
(\lambda_i(\cdot)), R_s)$ of an oligopoly pricing game is upper bounded by the number of relays $N$. The upper bound is
achieved when the cost derivatives are linear.
\end{theorem}

\textit{Proof:} Let the socially optimal routing be
$(r_i^*)_{i=1}^N=(\alpha_i R_s)_{i=1}^N$ where the coefficients
$(\alpha_i)$ are nonnegative and sum to one. The optimal cost then
is

\[
D^* = \sum_{i=1}^N \int_0^{\alpha_i R_s} \lambda_i(r)~dr.
\]

Since $\lambda_i(r)$ is concave, it can be shown that $\int_0^{\alpha_i R_s} \lambda_i(r)dr \ge \alpha_i^2 \int_0^{R_s}
\lambda_i(r)dr$ where equality holds when $\lambda_i(r)$ is linear. Therefore, $D^*$ is lower bounded as

\[
D^* \ge \sum_{i=1}^N \alpha_i^2 \int_0^{R_s} \lambda_i(r)~dr.
\]

Recall that inefficient equilibria in an oligopoly are monopolistic
such that the dominant relay $m$ satisfies
Theorem~\ref{thm:OligMonoEqui}. The price of anarchy, which is the
ratio of the cost at any monopolistic equilibrium (ME) to $D^*$, is
upper bounded as
\[
\frac{D^{ME}}{D^*} \stackrel{(a)}{\le}
\frac{\int_0^{R_s} \lambda_m(r)~dr}{\sum_{i=1}^N \alpha_i^2 \int_0^{R_s} \lambda_i(r)~dr}
\stackrel{(b)}{\le} \frac{\int_0^{R_s} \lambda_m(r)~dr}{\sum_{i=1}^N \alpha_i^2 \int_0^{R_s} \lambda_m(r)~dr}
= \frac{1}{\sum_{i=1}^N \alpha_i^2} \stackrel{(c)}{\le} N. \]

Next we specify the condition under which the upper bound is achieved. Notice that (a) holds with equality if and only
if $\int_0^{\alpha_i R_s} \lambda_i(r)dr = \alpha_i^2 \int_0^{R_s} \lambda_i(r)dr$ for all $i$. This requires each
$\lambda_i(r)$ to be a linear function. Inequality (b) is tight when $\int_0^{R_s} \lambda_i(r)dr = \int_0^{R_s}
\lambda_m(r)dr$ for every $i \ne m$. Hence, all the relays must have the same linear $\lambda_i(r)$.  Thus, $(r_i^*)$
must be the uniform, hence competitive, allocation, i.e., $\alpha_i = 1 / N$ for all $i$. This is exactly what is
needed to make (c) tight.

Now it remains to find the pricing functions which can induce the monopolistic equilibrium attaining the upper bound.
Let $\beta_i(r) = \lambda(R_s - r) \triangleq \beta(r)$ for every $i$.\footnote{Here we have omitted the subscript of
$\lambda_i(\cdot)$ in light of the symmetry.} Since $\beta_i(r)$ is strictly decreasing for every $i$, $\beta_{\hat
i}(r) = \beta(r) = \lambda(R_s - r)$. Since $\beta_{\hat i}(R_s - r) = \lambda(r)$, every relay is indifferent to
having any amount of flow. Thus, the monopolistic equilibrium can be sustained. \qed

Unlike the selfish routing games considered in~\cite{paper:Rou02,paper:Rou05}, for which the price of anarchy is
independent of the topology~\cite{paper:Rou03}, Theorem~\ref{thm:ConcavePOA} indicates that $\rho(N,
(\lambda_i(\cdot)), R_s)$ of an oligopoly PG explicitly depends on topology through $N$. Such a conclusion implies that
the more intensive (larger $N$) the competition is, the more inefficient the market becomes if it is monopolized. The
situation is even worse if the relays in an oligopoly have convex $\lambda_i(\cdot)$. In this case, the price of
anarchy can be arbitrarily large.

\vspace{0.1in}\begin{theorem}\label{thm:ConvexPOA} For a fixed
number $N \ge 2$ of relays and for any $M
> 0$, there exists an oligopoly $(N, (\lambda_{i}(\cdot))_{i=1}^N, R_s)$ with convex $(\lambda_{i}(\cdot))$ such that $\rho(N, (\lambda_{i}(\cdot)), R_s)
\ge M$.
\end{theorem}\vspace{0.1in}

{\textit{Sketch of Proof:} We can construct an oligopoly with $N$ relays such that the socially
optimal routing is competitive. By Corollary~\ref{cor:OligInefficient}, inefficient monopolistic
equilibria exist. However, within the class of convex functions, $\lambda_m(\cdot)$ of the dominant
relay $m$ can be designed so that $\int_0^{R_s}\lambda_m(\cdot) \ge M D^*$, where $D^*$ is the
optimal cost. \qed}

\subsection{Focal Equilibria}\label{subsec:OligFocalEqui}

Although possible, inefficient equilibria in an oligopoly are very unlikely to happen. The example
in Figure~\ref{fig:BadDuopoly} represents a highly pathological situation. Such an equilibrium is
reached only if the subtle relationships between $\beta_2(\cdot)$ and $\lambda_1(\cdot)$ and
between $\beta_1(\cdot)$ and $\lambda_2(\cdot)$ are satisfied. These relationships, however, can be
established only by coincidence, since relay $2$ cannot observe $\lambda_1(\cdot)$ and
relay $1$ cannot observe $\lambda_2(\cdot)$. In a general PG, it is arguably most rational for a relay to
use a replicating response as described in Section~\ref{subsec:OligBestResponse}.

\vspace{0.1in}\begin{definition}\label{def:FocalEqui} A focal equilibrium of a general pricing game
is an equilibrium where every relay $i$ adopts the replicating response to its local information
$\bs L_i = ((r_h, (B_j^h(\cdot))_{j \in {\cal S}_i^h})_{h \in {\cal P}_i}, (B_k^i(\cdot))_{k \in
{\cal O}_i})$, i.e., for all $h \in {\cal P}_i$, \be\label{eq:RepResponse1} B_i^h(t) \left\{\ba{ll}
= B_{\hat
i}^h(r_h) - B_{\hat i}^h(r_h - t), \quad&t \in [0, f_{hi}^*]\\
\ge B_{\hat i}^h(r_h) - B_{\hat i}^h(r_h - t),&t \in (f_{hi}^*, r_h]
\ea\right., \ee where $(B_{\hat i}^h(\cdot), f_{hi}^*)_{h \in {\cal
P}_i}$ are as specified in Lemma~\ref{lma:BestResp}.\footnote{Since
the derivative $\beta_{\hat i}^h(\cdot)$ of $B_{\hat i}^h(\cdot)$ is
in general piecewise continuous, we henceforth allow the derivative
$\beta_i^h(\cdot)$ of $B_{i}^h(\cdot)$ to be piecewise continuous.
Let $\beta_i^h(r)^-$ and $\beta_i^h(r)^+$ denote the left and right
limits of $\beta_i^h(\cdot)$ at $r$.}
\end{definition}\vspace{0.1in}

In this section, we investigate focal equilibria in oligopolies. Such equilibria are not only reasonable for
implementation, but also, more importantly, \emph{always} efficient. The next theorem establishes the existence of
focal equilibria in an oligopoly.

\vspace{0.1in}\begin{theorem}\label{thm:FocalEquiExist} The socially
optimal routing of an oligopoly is always induced by a focal
equilibrium.
\end{theorem}

\textit{Proof:} Note that the equilibrium constructed in the proof of Theorem~\ref{thm:OligOpt} is a focal equilibrium.
\qed

Figure~\ref{fig:DuopolyFocalEqui} illustrates a focal equilibrium in a duopoly induced by nonlinear pricing functions
$B_1(\cdot), B_2(\cdot)$. The linear pricing equilibrium used in the proof is a special case of
Figure~\ref{fig:DuopolyFocalEqui} such that the curves of $\beta_1(r)$ and $\beta_2(r)$ are the same horizontal line
that passes the point where $\lambda_1(r)$ and $\lambda_2(R_s - r)$ intersect. Notice that the focal equilibria
encompassed by the above example have a common property. That is, the curves $\beta_i(\cdot)$ and $\lambda_i(\cdot)$ of
all $i$ intersect at the point that corresponds to the social optimum. The next theorem states that such a phenomenon
is no coincidence. \vspace{0.1in}\begin{theorem}\label{thm:FocalEquiEff} Every focal equilibrium of an oligopoly is
efficient.
\end{theorem}\vspace{0.1in}

{\textit{Proof:} In light of Theorem~\ref{thm:OligCompEqui}, we only need to show that any
monopolistic focal equilibrium is efficient. Let $(\beta_i(\cdot))$ be the pricing functions that
induce such an equilibrium where $m$ is the dominant relay. By the best response
condition~\eqref{eq:OligBestResponse2} and the fact that $\beta_m(\cdot)$ is a replicating response
to $\beta_{\hat m}(\cdot)$, at $f_m^* = R_s$,
\[ \lambda_m(R_s) \le \beta_{\hat m}(0)^+ = \beta_m(R_s)^-.
\]
Applying \eqref{eq:OligBestResponse2} and
Lemma~\ref{lma:OligVirtualMargin} to any $j\ne m$, we have
\[
\lambda_j(0) \ge \beta_{\hat j}(R_s)^- \ge \beta_m(R_s)^-.
\]
Therefore, $\lambda_m(R_s) \le \lambda_j(0)$ for any $j \ne m$, which implies that the monopolistic routing is
efficient. \qed}

To summarize, as the most reasonable outcomes of a pricing game, focal equilibria always exist and are always efficient
in oligopolies. We have yet to find out whether these properties hold in general PGs. In the remainder of the paper, we
focus on the class of focal equilibria when we study pricing games in multi-hop networks.\footnote{We deliberately
ignore the type of inefficient equilibria discussed in Section~\ref{subsec:IneffiEqui} because there is no new
discovery we can make about them in the general PG. They are inefficient in oligopolies, and therefore inefficient in
general PGs.} For brevity, we will drop the qualifier ``focal'' henceforth.

\section{Equilibria in General Pricing Game}\label{sec:GeneralPG}

In this section, we consider a general multi-hop relay network with one source-destination pair as described in
Section~\ref{sec:Model}. As in Section~\ref{sec:Oligopoly}, we assume that in every local competition, a relay $i$
declares $\beta_i^h(\cdot) = p_i^h(\cdot) + d_{hi}(\cdot)$ to an $h\in{\cal P}_i$ and all $j \in {\cal S}_i^h$.

Notice that if $h\in{\cal P}_i$ has $r_h=0$, then technically any
$\beta_i^h(\cdot)$ is a best response since the local competition
involving $i$ and $j \in {\cal S}_i^h$ is vacuous. To prevent absurd
equilibria resulting from such arbitrary pricing, however, we assume
that $i$ uses \emph{honest pricing}
$\beta_i^h(t)=d_{hi}(t)+d_i(t+\sum_{h' \in {\cal P}_i \backslash
h}f_{h'i}^*)$ when $r_h=0$. Here, $f_{h'i}^*$ is the flow $i$
intends to acquire from $h' \in {\cal P}_i \backslash
h$,\footnote{Node $i$ need not use honest pricing for $h' \ne h$ if
$h'$ has positive incoming flow.} $d_i(t+\sum_{h' \in {\cal P}_i
\backslash h}f_{h'i}^*)$ is the derivative of the minimum cost
incurred by $i$ in forwarding traffic to its offsprings. Thus,
$\beta_i^h(t)$ exactly matches the actual cost of $i$ for forwarding
flow from $h$. Honest pricing, though restrictive, is in line with
$i$'s self interest. Being honest with its own cost, $i$ maximally
alleviates the burden of $h$, whose cost is partly leveraged by
$\beta_i^h(\cdot)$. Therefore, honest pricing can be seen as the
best effort by $i$ to help improve the competitiveness of $h$, in
the hope of earning profit from $h$ should $h$ later receive
positive $r_h$ from its predecessors.

We will frequently use the following terms. A \emph{path} is a concatenation of links from $s$ to $w$, while a
\emph{sub-path} is a contiguous segment of a path from a relay to $w$. Given a routing $(f_{ij})_{(i,j)\in{\cal E}}$, a
path/sub-path is said to have positive flow if $f_{mn}>0$ for every $(m,n)$ in that path/sub-path. Otherwise, the
path/sub-path is said to have zero flow. The \emph{marginal cost} on a path/sub-path is the sum of $d_{mn}(f_{mn})$
over all $(m,n)$ on that path/sub-path.

\vspace{0.1in}\begin{theorem}\label{thm:GeneralPGEqui} The socially optimal routing of a general PG
can be induced by an equilibrium.
\end{theorem}\vspace{0.1in}

Due to its length, the proof of the theorem is deferred to
Appendix~\ref{app:GeneralPGopt}.


\subsection{Inefficient Equilibria and Price of Anarchy}

\begin{figure}[!h]
  \begin{center}
  \includegraphics[width=7cm]{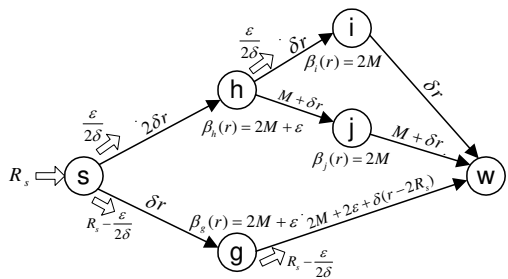}\\
  \caption{Arbitrarily bad equilibrium of a general PG.}\label{fig:MyopicEqui}
  \end{center}
\end{figure}

\begin{figure}[!h]
  \begin{center}
  \includegraphics[width=6cm]{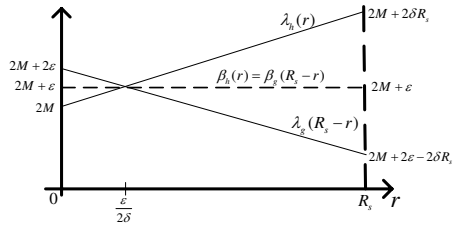}\\
  \caption{Competition between $h$ and $g$.}\label{fig:MyopicEqui1}
  \end{center}
\end{figure}

Unlike the oligopoly case, in a general PG, not all focal equilibria
are efficient.  The inefficiency of an equilibrium in general PG's
is caused not only by the manipulative behavior of dominant relays
but also by the {\em myopia} of nodes. We illustrate this point by
the game depicted in Figure~\ref{fig:MyopicEqui}.
The derivative of link cost functions are marked above each link, e.g. $d_{gw}(r)=2M+2\varepsilon+\delta(r-2R_s)$,
where $M$, $\varepsilon$ and $\delta$ are positive constants such that $M \gg \varepsilon R_s$ and $M \gg \delta R_s$.
The pricing function of each node is marked above the node. There are three paths from $s$ to $w$, of which $(s,h,i,w)$
has the smallest marginal cost $4\delta r$ even when $r=R_s$. So the socially optimal routing should allocate $R_s$
entirely to the path $(s,h,i,w)$. However, the equilibrium shown in Figure~\ref{fig:MyopicEqui} leads to only
$\varepsilon/(2\delta)$ being routed on $(s,h,i,w)$ while the rest is routed on $(s,g,w)$. In fact, $s$ is indifferent
among all allocations of $R_s$ to $h$ and $g$ since $\beta_h(\cdot) = \beta_g(\cdot) \equiv 2M+\varepsilon$.
Figure~\ref{fig:MyopicEqui1} explains why such $\beta_h(\cdot)$ and $\beta_g(\cdot)$ are $h$ and $g$'s best responses
in their competition.
Notice that $h$ is able to win only $\varepsilon/(2\delta)$ of the
total flow because it has cost $\lambda_h(r)=d_{sh}(r)+d_h(r) = 2M +
2\delta r$. This inflated cost is a consequence of $i$'s myopic
pricing. Since $i$ has superior cost ($\lambda_i(r)=2\delta r$)
relative to $j$ ($\lambda_j(r)=2M + 2\delta r$), $i$ can afford to
match $j$'s pricing function $\beta_j(\cdot)\equiv 2M$. Neither $i$
nor $j$ has any incentive to deviate from $2M$ since $i$ has
acquired all the flow $\varepsilon/(2\delta)$ while making the
maximum possible profit while $j$ would suffer a loss if it tried to
win a positive share by bidding lower than $2M$. Although $i$ could
have made more profit if it cut its price, thereby making $h$ more
competitive, it is unable to discover this opportunity as it lacks
``global vision''.

To conclude, although focal equilibria of a general PG rule out the manipulative pricing by a
superior relay (cf.~Sec.\ref{subsec:IneffiEqui}), the equilibria are susceptible to the
inefficiency caused by the dominant relays' myopic pricing. Such a source of inefficiency is
intrinsic to networks consisting of selfish nodes who are aware of their neighbors only. The price
of anarchy caused by myopic inefficiency can be arbitrarily large. For the example in
Figure~\ref{fig:MyopicEqui}, the equilibrium holds for all large enough $M$ and results in a total
cost
\[ D^E = \int_0^{\frac{\varepsilon}{2\delta}} 4 \delta
r ~dr + \int_0^{R_s
- \frac{\varepsilon}{2\delta}} 2(M+\varepsilon-\delta R_s) + 2 \delta r~dr = \frac{\varepsilon^2}{2\delta} + 2(M+\varepsilon-\delta R_s)\left(R_s -
\frac{\varepsilon}{2\delta}\right) + \delta \left(R_s -
\frac{\varepsilon}{2\delta}\right)^2, \]
whereas the optimal cost is
\[
D^* = \int_0^{R_s} 4\delta r~dr = 2 \delta R_s^2.
\]
Therefore, the price of anarchy is at least $D^E/D^*$, which can be made arbitrarily
large by increasing $M$.

Although in a multi-hop network equilibria can be arbitrarily inefficient, we will show in the
following that there is a class of equilibria which are always efficient.

\subsection{Everywhere Competitive Equilibria}

\begin{definition}
An equilibrium of a general PG is everywhere competitive if it induces a routing
$(f_{ij})_{(i,j)\in{\cal E}}$ such that $f_{hi}>0$ for at least two $i \in {\cal O}_h$
whenever $r_h>0$ unless $w \in {\cal O}_h$ and $f_{hw}>0$.\footnote{Recall that we have
assumed in Section~\ref{sec:network} that either ${\cal O}_h$ contains at least two
relays or ${\cal O}_h = \{w\}$.}
\end{definition}\vspace{0.1in}

Notice that the equilibrium in Figure~\ref{fig:MyopicEqui} is not everywhere competitive
as $i$ is dominant to $j$. One would expect that when no relay is unrivalled, mistakes
such as the one made by $i$ could be avoided. The next theorem validates this intuition.
Its proof is contained in the Appendix~\ref{app:EverywhereComp}.

\begin{theorem}\label{thm:EverywhereComp} If an equilibrium of a general PG is
everywhere competitive, it must be efficient.
\end{theorem}\vspace{0.1in}


\section{Pricing Game with Elastic Source}\label{sec:ElasticPG}

So far we have assumed that $s$ has a fixed, inelastic demand. In
this section, we show that pricing games with an elastic source can
be studied within the same framework we have developed for the
inelastic case.

We consider a source $s$ with elastic traffic demand. The source's preference over different admitted rates $r_s$ is
measured by a utility function $U_s(r_s)$ such that $U_s(r_s) = U_s(R_s)$ for all $r_s \ge R_s$. In other words, $R_s$
is the maximum desired service rate of $s$. In the interval $[0, R_s]$, $U_s(\cdot)$ is assumed to be strictly
increasing, concave with continuous derivative $u_s(\cdot)$. Taking the approach of~\cite{book:BG92}, we define the
{\em overflow rate} as $f_{sw} \triangleq R_s - r_s$. Thus, at $s$ we have
\begin{equation}\label{eq:SourceFlowCons}
\sum_{i \in {\cal O}_s} {f_{si}} + f_{sw} = R_s.
\end{equation}

\begin{figure}[!h]
  \begin{center}
  \includegraphics[width=5cm]{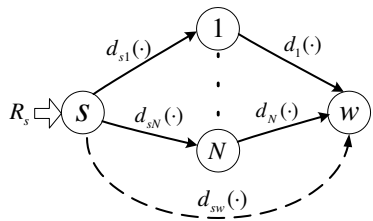}\\
  \caption{Oligopoly with an overflow link.}\label{fig:Overflow}
  \end{center}
\end{figure}

Let $D_{sw}(f_{sw}) \triangleq U_s(R_s) - U_s(r_s)$ denote the utility loss to $s$ resulting from
having a rate of $f_{sw}$ rejected from the network. Equivalently, if we imagine that the blocked
flow $f_{sw}$ is routed on a {\em virtual overflow link} directly from $s$ to $w$ \cite{book:BG92},
then $D_{sw}(f_{sw})$ can simply be interpreted as the cost incurred on the overflow link when its
flow rate is $f_{sw}$. Moreover, as defined, $D_{sw}(f_{sw})$ is strictly increasing, continuously
differentiable, and convex in $f_{sw}$ on $[0, R_s]$. Denote the derivative of $D_{sw}(\cdot)$ by
$d_{sw}(\cdot)$. Thus, we can treat the pricing game with an elastic source as one with an inelastic
source and an overflow link $(s,w)$. An oligopoly pricing game with an overflow link is
illustrated in Figure \ref{fig:Overflow}, where the overflow link $(s,w)$ is represented by a
dashed arrow.
Such an oligopoly is essentially the same as those studied in Section~\ref{sec:Oligopoly} with the exception that $s$
now has the additional option of sending traffic on link $(s,w)$. From a pricing perspective, we can think of $w$ as
directly competing with relays by using a \emph{uniformly-zero pricing function}.

In a general pricing game, the introduction of the overflow link affects only the local competition
faced by $i \in {\cal O}_s$. Now $w$ becomes a new competitor to all $i \in {\cal O}_s$ whose
presence changes each $i$'s perception of the competition. Specifically, the pricing function of
$i$'s virtual competitor is derived as
\[
B_{\hat i}^s(r) = \min \sum_{j \in {\cal S}_i^s} B_j^h(f_{sj}) +
D_{sw}(f_{sw}),
\]
where the minimization is taken with respect to nonnegative $((f_{sj})_{j \in {\cal S}_i^s}, f_{sw})$ such that $\sum_j
f_{sj} + f_{sw} = r$. The conclusions for pricing games with an elastic source are almost verbatim to those for
inelastic pricing games. Limited by space, we do not elaborate further.

\section{Conclusion}

This work presented a game-theoretic analysis of price competition
in multi-hop relay networks. The introduction of possibly nonlinear
pricing functions to the game enabled us to develop a much richer
set of results than if we allowed only constant unit prices. While
the socially optimal routing can always be induced by an
equilibrium, the game may have inefficient equilibria as well.
Furthermore, the existence of competition turns out to be a
two-sided coin.  On the one side, any competitive equilibrium in
oligopoly pricing games and any everywhere competitive equilibrium
in general pricing games must be efficient. On the other side, the
conclusion that the price of anarchy of an oligopoly is equal to the
number of competitors seems to suggest that more intense competition
only makes inefficient (monopolistic) equilibria even worse. Unlike
the case of oligopolies, the inefficiency of equilibria in a general
pricing game can be attributed not only to manipulative pricing by
dominant relays, but also more fundamentally, to the myopia of
dominant relays.  We showed that the price of anarchy attributed to
both the monopolistic and myopic effects is unbounded.

\appendix

\subsection{Proof of Theorem~\ref{thm:GeneralPGEqui}}\label{app:GeneralPGopt}

We prove the theorem by constructing an equilibrium that supports the socially optimal
routing. Let $(f_{hi}^*)$ be the link flows of the socially optimal routing. Denote the
total incoming flow at node $i$ by $r_i^*\triangleq\sum_{h\in{\cal P}_i} f_{hi}^*$. Note
that under the socially optimal routing, a path/sub-path has positive flow only if it has
the minimum marginal cost among all paths/sub-paths with the same origin. Let
$\lambda_i^*$ denote the minimum marginal cost of any path/sub-path from $i$ to $w$ (for
$i=w$, $\lambda_i^* = 0$). Consider the following pricing scheme. Each relay $i$ adopts
$\beta_i^h(t)\equiv \lambda_h^*$ for any $h\in{\cal P}_i$ unless $r_h^*=0$, in which case
honest pricing $\beta_i^h(t) = d_{hi}(t) + d_i(r_i^*+t)$ is enforced. Such a pricing
scheme will be referred to as \emph{marginal cost pricing}. We show that marginal cost
pricing supports the socially optimal routing. That is, for any relay $i$, given that
each $h\in {\cal P}_i$ has total traffic $r_h^*$ to allocate\footnote{We consider only
those $h\in {\cal P}_i$ with $r_h^*>0$, since any predecessor with zero traffic is
irrelevant to the determination of $i$'s best response, and so can be ignored.} and every
relay $j \in {\cal S}_i^h$ adopts $\beta_j^h(t)\equiv \lambda_h^*$ (if $w\in {\cal
S}_i^h$, it always adopts $\beta_w^h(t) = d_{hw}(t)$), (1) $\beta_i^h(t)\equiv
\lambda_h^*$ for each $h \in {\cal P}_i$ is a best response and (2) the traffic
allocation $(f_{hi}^*)_{h\in{\cal P}_i}$ is the most profitable allocation from $i$'s
perspective (cf.~Lemma~\ref{lma:BestResp}).

We will need the following lemmas to prove the theorem. \vspace{0.1in}

\begin{lemma}\label{lma:IncreasingMargCost}
Under marginal cost pricing, for each relay $i$, its virtual competitor's pricing
function $B_{\hat i}^h(\cdot)$ and its minimum outgoing cost function $D_i(\cdot)$ are
convex.
\end{lemma}
\vspace{0.1in}

Lemma~\ref{lma:IncreasingMargCost} is a special case of the next general observation.

\vspace{0.1in}
\begin{lemma}\label{lma:GeneralConvex}
For any $K\in\mathbb{N}$, the function
\[
g(x) \triangleq \min_{\substack{\sum_{i=1}^K x_i = x\\x_i\in{\cal G}_i,i=1,\cdots,K}}
\sum_{i=1}^K g_i(x_i)
\]
is convex if every $g_i(\cdot)$ with domain ${\cal G}_i$ is convex.
\end{lemma}
\textit{Proof:} Let $y, z$ be two distinct points in the domain of $g(\cdot)$. For any
$\lambda\in[0,1]$, let $\bar\lambda$ denote $1-\lambda$. We have \beas \lambda g(y) +
\bar\lambda g(z) &=& \lambda \min_{\substack{\sum_i y_i = y\\y_i\in{\cal G}_i}} \sum_i
g_i(y_i) + \bar\lambda
\min_{\substack{\sum_i z_i = z\\z_i\in{\cal G}_i}} \sum_i g_i(z_i)\\
&\stackrel{(a)}{=}& \lambda\sum_i g_i(y_i^*) + \bar\lambda\sum_i g_i(z_i^*)\\
&\stackrel{(b)}{\ge}& \sum_i  g_i(\lambda y_i^*+\bar\lambda z_i^*) \\
&\stackrel{(c)}{\ge}& \min_{\substack{\sum_i x_i = \lambda y + \bar\lambda z\\x_i\in{\cal
G}_i}} \sum_i g_i(x_i) = g(\lambda y + \bar\lambda z).  \eeas For equality (a), we assume
$(y_i^*) =\arg\min_{\sum_i y_i = y} \sum_i g_i(y_i)$ and $(z_i^*) =\arg\min_{\sum_i z_i =
z} \sum_i g_i(z_i)$. Inequality (b) follows from the convexity of each $g_i(\cdot)$.
Because $\sum_i \lambda y_i^*+\bar\lambda z_i^* = \lambda y + \bar\lambda z$ and $\lambda
y_i^*+\bar\lambda z_i^* \in {\cal G}_i$ for $i=1,\cdots,K$, we have inequality (c). \qed

\textit{Proof of Lemma~\ref{lma:IncreasingMargCost}:} By definition,
\[
B_{\hat i}^h(t) = \min_{\sum_{j\in{\cal S}_i^h} f_{hj} = t} \sum_{j\in{\cal S}_i^h}
B_j^h(f_{hj}),
\]
where $B_j^h(t) = \int_0^{t} \beta_j^h(r)~dr$. Under marginal cost pricing $\beta_j^h(r)
\equiv \lambda_h^*$ if $j$ is a relay or $\beta_j^h(r) = d_{hj}(r)$ if $j$ is the
destination, in either case $B_j^h(t)$ is convex. So $B_{\hat i}^h(t)$ is convex by
Lemma~\ref{lma:GeneralConvex}. Similarly, we have
\[
D_i(t) = \min_{\sum_{k\in{\cal O}_i} f_{ik} = t} \sum_{k\in{\cal O}_i} B_k^i(f_{ik}),
\]
which is convex also because each $B_k^i(t)$ is convex. \qed

Lemma~\ref{lma:IncreasingMargCost} implies that under marginal cost pricing, $\beta_{\hat
i}^h(t) = \frac{d}{dt}B_{\hat i}^h(t)$ and $d_i(t) = \frac{d}{dt}D_i(t)$ are
nondecreasing for all relays $i$ and all $h\in{\cal P}_i$. In fact, $\beta_{\hat i}^h(t)$
and $d_i(t)$ have the following characterization. \vspace{0.1in}
\begin{lemma}\label{lma:Characterization}
Under marginal cost pricing, $\beta_{\hat i}^h(t)$ of any relay $i$ and any $h\in{\cal
P}_i$ with positive incoming traffic can be expressed as\footnote{Recall the assumption
that either ${\cal O}_h$ contains at least two relays or ${\cal O}_h = \{w\}$.}
\be\label{eq:BetaHatI} \beta_{\hat i}^h(t) = \left\{\ba{ll}
    \lambda_h^*, \quad& \textrm{if}~w \notin{\cal S}_i^h\\
    \min(d_{hw}(t),\lambda^*_h), & \textrm{otherwise}
     \ea \right.,
\ee and if $i$ has positive incoming traffic, then $d_i(t)$ can be expressed as
\be\label{eq:Di} d_{i}(t) = \left\{\ba{ll}
    \lambda_i^*, \quad& \textrm{if}~w \notin{\cal O}_i\\
    d_{iw}(t), & \textrm{if}~{\cal O}_i = \{w\}\\
    \min(d_{iw}(t),\lambda^*_i), & \textrm{otherwise}
     \ea \right..
     \ee
\end{lemma}
\textit{Proof:} First, by assumption, ${\cal S}_i^h$ contains at least one relay and
every relay $j$ in ${\cal S}_i^h$ applies $\beta_j^h(t) \equiv \lambda_h^*$. If $w \notin
{\cal S}_i^h$, the (marginal) pricing function of $i$'s virtual competitor is nothing but
$\beta_{\hat i}^h(t) \equiv \lambda_h^*$. However, if $w \in {\cal S}_i^h$, it always
adopts $\beta_w^h(t) = d_{hw}(t)$. Thus, the pricing function of $i$'s virtual competitor
is
\[
B_{\hat i}^h(t) = \min_{x+y=t}\int_0^x d_{hw}(r)~dr + \int_0^y \lambda_h^* ~dr.
\]
Because $d_{hw}(r)$ is increasing, $\beta_{\hat i}^h(t) = \frac{d}{dt}B_{\hat i}^h(t) =
\min(d_{hw}(t),\lambda_h^*)$. The derivation of~\eqref{eq:Di} is almost verbatim, and we
we skip it for brevity. \qed

While Lemma~\ref{lma:Characterization} provides general expressions for $\beta_{\hat
i}^h(t)$ and $d_i(t)$ under marginal cost pricing, the following lemma specifies the
values of the two functions evaluated at the optimal routing configuration, i.e.,
$\beta_{\hat i}^h(r_h^*-f_{hi}^*)$ and $d_i(r_i^*)$. These values are crucial to
determining a best response $\beta_i^h(\cdot)$ for all $h\in{\cal P}_i$. \vspace{0.1in}
\begin{lemma}\label{lma:CharacterizationEqui}
Under marginal cost pricing and at the socially optimal routing
$(f_{mn}^*)_{(m,n)\in{\cal E}}$ with $r_n^* = \sum_{m\in{\cal P}_n}f_{mn}^*$, $n\in{\cal
N}$, we have for any relay $i$ and any $h\in{\cal P}_i$, $\beta_{\hat
i}^h(r_h^*-f_{hi}^*) = \lambda_h^*$ and $d_i(r_i^*) = \lambda_i^*$. Furthermore,
$\beta_{\hat i}^h(t) \equiv \lambda_h^*$ for $t\in[r_h^*-f_{hi}^*, r_h^*]$.
\end{lemma}
\textit{Proof:} By \eqref{eq:BetaHatI}, $\beta_{\hat i}^h(r_h^*-f_{hi}^*)$ can be
evaluated as follows:
\[
\beta_{\hat i}^h(r_h^*-f_{hi}^*) = \left\{\ba{ll}
    \lambda_h^*, \quad& \textrm{if}~w \notin{\cal S}_i^h\\
    \min(d_{hw}(r_h^*-f_{hi}^*),\lambda^*_h) = \lambda^*_h, & \textrm{otherwise}
     \ea \right.,
\]
where the second equality holds since $d_{hw}(r_h^*-f_{hi}^*) \ge d_{hw}(f_{hw}^*)
\ge\lambda_h^*$. So in summary, $\beta_{\hat i}^h(r_h^*-f_{hi}^*) = \lambda^*_h$ under
all circumstances. Moreover, \eqref{eq:BetaHatI} also implies that $\beta_{\hat
i}^h(t)\le \lambda_h^*$ for all $t\in[0, r_h^*]$. However, $\beta_{\hat i}^h(t)$ is
nondecreasing as a result of Lemma~\ref{lma:IncreasingMargCost}. Thus, $\beta_{\hat
i}^h(t) \equiv \lambda_h^*$ for $t\in[r_h^*-f_{hi}^*, r_h^*]$.

Next we show that under marginal cost pricing, $d_i(r_i^*) = \lambda_i^*$ for all nodes
$i$. Let's first assume $r_i^*>0$ and use the expression~\eqref{eq:Di}. If ${\cal O}_i =
\{w\}$, $d_i(t) = \beta_w^i(t) = d_{iw}(t)$. In this case, $f_{iw}^* = r_i^*$ and so
$d_i(r_i^*) = d_{iw}(f_{iw}^*) = \lambda_i^*$. If $w\notin {\cal O}_i$, then all the
offsprings of $i$ announce $\beta_j^i(\cdot)\equiv \lambda_i^*$, thus $d_i(\cdot)\equiv
\lambda_i^*$. Finally, if both $w$ and some relay $j$ are in ${\cal O}_i$, then $d_i(t) =
\min(d_{iw}(t), \lambda_i^*)$. Because $d_{iw}(r_i^*) \ge d_{iw}(f_{iw}^*)
\ge\lambda_i^*$, we have $d_i(r_i^*) = \min(d_{iw}(r_i^*), \lambda_i^*) = \lambda_i^*$.
So we are done with the case $r_i^*>0$.

If $r_i^*=0$, every offspring $j$ of $i$ adopts honest pricing $\beta_j^i(t) = d_{ij}(t)
+ d_j(r_j^* + t)$ (and $\beta_w^i(t) = d_{iw}(t)$ if $w\in {\cal O}_i$). Thus, by
defining $d_w(\cdot)\equiv 0$, we can express $d_i(r_i^*)$ as
\[
d_i(r_i^*) = d_i(0) = \min_{j\in{\cal O}_i} \beta_j^i(0) = \min_{j\in{\cal O}_i}
d_{ij}(0) + d_j(r_j^*).
\]
If $d_j(r_j^*) = \lambda_j^*$ for all $j \in {\cal O}_i$, $d_i(r_i^*)$ must be equal to
$\lambda_i^*$. So far we have shown that $d_j(r_j^*) = \lambda_j^*$ if $r_j^* > 0$.
Should $r_j^*=0$ for some $j \in {\cal O}_i$, we can evaluate $d_j(r_j^*)$ by $d_j(r_j^*)
= d_j(0) = \min_{k\in{\cal O}_j} d_{jk}(0) + d_k(r_k^*)$. If $r_k^*> 0$ for all
$k\in{\cal O}_j$, then we will be able to show that $d_j(r_j^*) = \lambda_j^*$.
Otherwise, we further expand those $d_k(r_k^*)$ such that $r_k^*= 0$. Because the routing
topology is loop-free, eventually the recursive evaluation will terminate (either when
all offsprings have positive incoming traffic or when the destination is the only
offspring\footnote{If $w$ is the only offspring of a relay, say $i$, then $r_i^* =
f_{iw}^*$ and $d_i(r_i^*) = d_{iw}(f_{iw}^*) =\lambda_i^*$.}). Then calculating in the
reverse order of the recursive expansion, we can in the end show that $d_i(r_i^*) =
\lambda_i^*$ even if $r_i^*=0$. So the proof is complete. \qed

Now we are ready to apply Lemma~\ref{lma:BestResp} to show that (1) using
$\beta_i^h(t)\equiv \lambda_h^*$ for all $h\in{\cal P}_i$ is a best response of $i$ given
each $h\in{\cal P}_i$ has total traffic $r_h^*$ and its competitors all adopt marginal
cost pricing and (2) this best response induces traffic allocation $(f_{hi}^*)_{h\in{\cal
P}_i}$.

\textit{Proof of Theorem~\ref{thm:GeneralPGEqui}:}

\textbf{Step 1:} Referring to Lemma~\ref{lma:BestResp}, we first show that
$(f_{hi}^*)_{h\in{\cal P}_i}$ maximizes \beas \bar\Gamma_i(\bs f_i; \bs L_i) &=& \sum_{h
\in {\cal P}_i} \left[B_{\hat i}^h(r_h^*) - B_{\hat i}^h(r_h^* - f_{hi})
- D_{hi}(f_{hi}) \right] - D_i\left(\sum_{h \in {\cal P}_i}f_{hi}\right)\\
&=& \sum_{h \in {\cal P}_i} \int_0^{f_{hi}} \beta_{\hat i}^h(r_h^* - t) - d_{hi}(t)~dt -
\int_0^{\sum_{h \in {\cal P}_i}f_{hi}} d_i(t)~dt \eeas over all the $\bs f_i = (f_{hi})$
such that $0 \le f_{hi} \le r_h^*$ for all $h \in {\cal P}_i$.

Note that if $f_{hi}^* > 0$, $i$ must be on a sub-path from $h$ to $w$ with the minimum
marginal cost, so $\lambda_h^* = d_{hi}(f_{hi}^*) + \lambda_i^*$. Otherwise, $\lambda_h^*
\le d_{hi}(f_{hi}^*) + \lambda_i^*$. Therefore, we can conclude that \beas
(f_{hi}^*)_{h\in{\cal P}_i} &=& \mathop{\arg\max}_{\substack{0 \le f_{hi}\le
r_h^*\\h\in{\cal P}_i}} \left\{ \sum_{h\in{\cal P}_i} \int_0^{f_{hi}} \beta_{\hat
i}^h(r_h^* - t) - d_{hi}(t) - d_i(r_i^* - f_{hi}^* + t)~dt
\right\}\\
&=& \mathop{\arg\max}_{\substack{0 \le f_{hi}\le r_h^*\\h\in{\cal P}_i}}\left\{
\bar\Gamma_i(\bs f_i; \bs L_i) + \int_0^{\sum_{h\in{\cal P}_i} f_{hi}}
d_i(t)~dt - \sum_{h\in{\cal P}_i} \int_0^{f_{hi}} d_i(r_i^* - f_{hi}^* + t)~dt \right\}\\
&\triangleq& \mathop{\arg\max}_{\substack{0 \le f_{hi}\le r_h^*\\h\in{\cal P}_i}}\left\{
\bar\Gamma_i(\bs f_i; \bs L_i) + \Delta(\bs f_i) \right\}. \eeas The first equality
follows from the fact that $\beta_{\hat i}^h(r_h^* - t) - d_{hi}(t) - d_i(r_i^* -
f_{hi}^* + t)$ is nonincreasing in $t$ (by Lemma~\ref{lma:IncreasingMargCost}) and that
\be\label{eq:CompSlackness} \beta_{\hat i}^h(r_h^* - f_{hi}^*) - d_{hi}(f_{hi}^*) -
d_i(r_i^*) \left\{\ba{ll}
    \le 0, \quad & f_{hi}^* = 0\\
    =0, & 0 < f_{hi}^* \le r_m^* \ea \right.,
\ee which follows from Lemma~\ref{lma:CharacterizationEqui}. The second and third
equalities are by regrouping terms and defining the terms other than $\bar\Gamma_i(\bs
f_i; \bs L_i)$ to be $\Delta(\bs f_i)$. We can rewrite $\Delta(\bs f_i)$ as
\[
\Delta(\bs f_i) = \int_0^{r_i} d_i(t)~dt - \sum_{h\in{\cal P}_i} \int_{r_i^* -
f_{hi}^*}^{r_i^* - f_{hi}^* + f_{hi}} d_i(t)~dt.
\]
Now consider the difference
\[
\Delta(\bs f_i) - \Delta(\bs f_i^*) = \int_{r_i^*}^{r_i}d_i(t)~dt - \sum_{h\in{\cal P}_i}
\int_{r_i^*}^{r_i^* - f_{hi}^* + f_{hi}} d_i(t)~dt.
\]
Recall that $d_i(\cdot)$ is positive and nondecreasing. If $f_{hi} \ge f_{hi}^*$ for all
$h\in{\cal P}_i$, then the difference is easily seen to be nonnegative. If $f_{hi} \le
f_{hi}^*$ for all $h\in{\cal P}_i$, we can rewrite the difference as
\[
\Delta(\bs f_i) - \Delta(\bs f_i^*) = \sum_{h\in{\cal P}_i} \int_{r_i^* - f_{hi}^* +
f_{hi}}^{r_i^*} d_i(t)~dt - \int_{r_i}^{r_i^*}d_i(t)~dt,
\]
which, by the same reason, is nonnegative.

To summarize, the function $\Delta(\bs f_i)$ is minimized at $\bs f_i^*$ within the
region ${\cal F}(\bs f_i^*) \triangleq \{\bs f_i: f_{hi} \ge f_{hi}^*, \forall h\in{\cal
P}_i~\textrm{or}~f_{hi} \le f_{hi}^*, \forall h\in{\cal P}_i\}$. Thus, it can be
concluded that $\bar\Gamma_i(\bs f_i; \bs L_i)$ must be maximized at $\bs f_i^*$ within
${\cal F}(\bs f_i^*)$. Recall that the feasible region for the maximization of
$\bar\Gamma_i(\bs f_i; \bs L_i)$ is ${\cal F}_i = \{\bs f_i: 0\le f_{hi} \le r_{h}^*,
\forall h\in{\cal P}_i\}$. Next we show that the maximizer of $\bar\Gamma_i(\bs f_i; \bs
L_i)$ cannot lie in ${\cal F}^c(\bs f_i^*) \triangleq {\cal F}_i \backslash {\cal F}(\bs
f_i^*)$.

Suppose $\bar\Gamma_i(\bs f_i; \bs L_i)$ is maximized at $\bar{\bs f_i}$. Then at the
maximum it must hold for all $h\in{\cal P}_i$ that
\[
\frac{\partial}{\partial f_{hi}} \bar\Gamma_i(\bar{\bs f_i};\bs L_i) = \beta_{\hat
i}^h(r_h^* - \bar f_{hi}) - d_{hi}(\bar f_{hi}) - d_i(\bar r_i) \left\{\ba{ll}
    \le 0, \quad& \bar f_{hi} = 0\\
    = 0, & 0 < \bar f_{hi} < r_h^*\\
    \ge 0, & \bar f_{hi} = r_h^* \ea \right.,
\]
or equivalently \be\label{eq:SimultaneousFormulas} \beta_{\hat i}^h(r_h^* - \bar f_{hi})
- d_{hi}(\bar f_{hi})\left\{\ba{ll}
    \le d_i(\bar r_i), \quad& \bar f_{hi} = 0\\
    = d_i(\bar r_i), & 0 < \bar f_{hi} < r_h^*\\
    \ge d_i(\bar r_i), & \bar f_{hi} = r_h^* \ea \right.,
\ee where $\bar r_i = \sum_h \bar f_{hi}$. It is straightforward to verify that $\bs
f_i^*$ is a solution to the above simultaneous formulas (cf~\eqref{eq:CompSlackness}).
Next we demonstrate that there is no other solution in ${\cal F}^c(\bs f_i^*)$. First, if
$f_{hi}^* = 0$ for all $h\in{\cal P}_i$ or if $f_{hi}^* = r_h^*$ for all $h\in{\cal
P}_i$, then ${\cal F}^c(\bs f_i^*)$ is empty, so we are done. Otherwise, let ${\cal
P}_i^0 \triangleq \{h\in{\cal P}_i: f_{hi}^*=0\}$, ${\cal P}_i^+ \triangleq \{h\in{\cal
P}_i: f_{hi}^*=r_h^*\}$. Hence, for those $h\in{\cal P}_i\backslash({\cal P}_i^0\cup{\cal
P}_i^+)$, $f_{hi}^*$ is in the interior of $[0, r_h^*]$. Accordingly, the simultaneous
formulas satisfied by $\bs f_i^*$ are \beas
\beta_{\hat i}^h(r_h^*) - d_{hi}(0) &\le& d_i(r_i^*), \quad h \in {\cal P}_i^0,\\
\beta_{\hat i}^h(r_h^* - f_{hi}^*) - d_{hi}(f_{hi}^*) &\ge& d_i(r_i^*), \quad h \in {\cal P}_i^+,\\
\beta_{\hat i}^h(0) - d_{hi}(r_h^*) &=& d_i(r_i^*), \quad \textrm{otherwise}. \eeas

Now suppose there is a different solution $\bs f_i'$ in ${\cal F}^c(\bs f_i^*)$ such that
for some $a,b\in{\cal P}_i$, $f_{ai}'
> f_{ai}^*$ and $f_{bi}' < f_{bi}^*$. It follows that $a\notin{\cal P}_i^+$ and $b \notin {\cal P}_i^0$. Hence, $\beta_{\hat i}^a(r_a^* - f_{ai}^*) - d_{ai}(f_{ai}^*) \le
\beta_{\hat i}^b(r_b^* - f_{bi}^*) - d_{bi}(f_{bi}^*)$. Because $\beta_{\hat i}^h(r_h^* -
f_{hi}) - d_{hi}(f_{hi})$ is strictly decreasing with $f_{hi}$ for all $h\in{\cal P}_i$,
we must have $\beta_{\hat i}^a(r_a^* - f_{ai}') - d_{ai}(f_{ai}') < \beta_{\hat
i}^b(r_b^* - f_{bi}') - d_{bi}(f_{bi}')$. However, since $f_{ai}' > 0$, $f_{bi}' <
r_b^*$, and they satisfy the respective formulas in~\eqref{eq:SimultaneousFormulas}, it
follows that
\[
\beta_{\hat i}^a(r_a^* - f_{ai}') - d_{ai}(f_{ai}') \ge d_i(r_i') \ge \beta_{\hat
i}^b(r_b^* - f_{bi}') - d_{bi}(f_{bi}'),
\]
thus, a contradiction. Therefore, we have shown that no solution exists in ${\cal
F}^c(\bs f_i^*)$.

To recap, the maximizer of $\bar\Gamma_i(\bs f_i; \bs L_i)$ must fall in ${\cal F}(\bs
f_i^*)$. Moreover, we have found that $\bs f_i^*$ maximizes $\bar\Gamma_i(\bs f_i; \bs
L_i)$ within ${\cal F}(\bs f_i^*)$. So we can conclude that
\[
\bs f_i^* = \mathop{\arg\max}_{{\cal F}_i}\bar\Gamma_i(\bs f_i; \bs L_i).
\]

\textbf{Step 2:} Next we show that condition~\eqref{eq:BestResponse1} holds with
$\beta_i^h(t)\equiv \lambda_h^*$, i.e, \[ B_i^h(t) \ge B_{\hat i}^h(r_h^*) - B_{\hat
i}^h(r_h^* - t)
\]
for all $h\in{\cal P}_i$ and all $t\in[0,r_h^*]$. Since $\beta_i^h(t)\equiv \lambda_h^*$,
LHS is $B_i^h(t) = \int_0^t \beta_i^h(r)~dr = \lambda_h^* t$. By the
expression~\eqref{eq:BetaHatI}, $\beta_{\hat i}^h(t) \le \lambda_h^*$ under all
circumstances. So the RHS of the above inequality is $B_{\hat i}^h(r_h^*) - B_{\hat
i}^h(r_h^* - t) = \int_{r_h^*-t}^{r_h^*} \beta_{\hat i}^h(r)~dr \le \lambda_h^* t$. And
this proves~\eqref{eq:BestResponse1}.

\textbf{Step 3:} Finally, we prove that condition~\eqref{eq:BestResponse2} is established
by the combination of the pricing function $\beta_i^h(t)\equiv \lambda_h^*$ and the
traffic allocation $f_{hi}^*$, i.e., for all $h\in{\cal P}_i$, \[ B_i^h(f_{hi}^*) =
B_{\hat i}^h(r_h^*) - B_{\hat i}^h(r_h^* - f_{hi}^*).
\]
Note that the LHS is equal to $B_i^h(f_{hi}^*) = \int_0^{f_{hi}^*} \beta_i^h(r)~dr =
\lambda_h^* f_{hi}^*$. By Lemma~\ref{lma:CharacterizationEqui}, $\beta_{\hat
i}^h(t)\equiv \lambda_h^*$ for $t\in[r_h^*-f_{hi}^*, r_h^*]$. It follows that $B_{\hat
i}^h(r_h^*) - B_{\hat i}^h(r_h^* - f_{hi}^*) = \int_{r_h^* - f_{hi}^*}^{r_h^*}\beta_{\hat
i}^h(r)~dr = \lambda_h^* f_{hi}^*$. So \eqref{eq:BestResponse2} is proved.

\textbf{Conclusion:} With all the necessary and sufficient conditions in
Lemma~\ref{lma:BestResp} satisfied, we can conclude that $\beta_i^h(t)\equiv \lambda_h^*$
for all $h\in{\cal P}_i$ is a best response of $i$ given that all its competitors also
adopt marginal cost pricing. Furthermore, the socially optimal routing $(f_{hi}^*)$ is
the traffic allocation induced by the equilibrium under marginal cost pricing. Therefore,
the proof for Theorem~\ref{thm:GeneralPGEqui} is complete. \qed

\subsection{Proof of Theorem~\ref{thm:EverywhereComp}}\label{app:EverywhereComp}

We use the following lemmas to prove the theorem.

\begin{lemma}\label{lma:EverywhereComp1}
Given any (focal) equilibrium with induced link flows $(f_{hi})$ and node total incoming
rates $(r_i)$, for any node $i$ with $r_i\in(0,R_s)$, its actual marginal cost
$\lambda_i^h(\cdot)$ of forwarding traffic for any $h\in{\cal P}_i$ (with $f_{h'i}$ of
all other $h' \in {\cal P}_i$ fixed) satisfies \be\label{eq:LambdaIneq}
\lambda_i^h(f_{hi})^+ \le \lambda_i^h(f_{hi})^-, \ee where $\lambda_i^h(f_{hi})^+$ and
$\lambda_i^h(f_{hi})^-$ are the right and left limits of $\lambda_i^h(\cdot)$ at
$f_{hi}$.
\end{lemma}

\textit{Proof:} First assume that $i$ has only one offspring, which by our assumption
must be the destination. Thus, $\lambda_i^h(t) = d_{hi}(t) + d_{iw}(t)$. Since both
$d_{hi}(\cdot)$ and $d_{iw}(\cdot)$ are continuous everywhere, $\lambda_i^h(\cdot)$ must
also be continuous everywhere. The inequality~\eqref{eq:LambdaIneq} holds with equality.

Next assume that $i$ has multiple offsprings. By the same reasoning as used in the proof
of Lemma~\ref{lma:OligVirtualMargin}, for any $j \in {\cal O}_i$ with $f_{ij}>0$, we have
\[
\lambda_i^h(f_{hi})^+ \le d_{hi}(f_{hi}) + \beta_{\hat j}^i(r_i-f_{ij})^+ =
d_{hi}(f_{hi}) + \beta_{j}^i(f_{ij})^- \le \lambda_i^h(f_{hi})^-,
\]
where the equality follows from the replicating response $\beta_{j}^i(t) = \beta_{\hat
j}^i(r_i-t), ~t\in[0,f_{ij}]$, assumed by a focal equilibrium. \qed

\begin{lemma}\label{lma:EverywhereComp2}
Given any (focal) equilibrium with induced link flows $(f_{hi})$ and node total incoming
rates $(r_i)$, for any $h \ne w$ such that $f_{hi}>0$ and $f_{hj}>0$ for two different
relays $i,j\in{\cal O}_h$, it holds that

(i) the actual marginal cost $\lambda_i^h(\cdot)$, $\lambda_j^h(\cdot)$ of $i$ and $j$
forwarding traffic for $h$ are continuous at $f_{hi}$ and $f_{hj}$, respectively;

(ii) the marginal pricing functions $\beta_{\hat i}^h(\cdot)$, $\beta_{\hat j}^h(\cdot)$
of $i$ and $j$'s virtual competitors are continuous at $r_h-f_{hi}$ and $r_h-f_{hj}$,
respectively;

(iii)
\[
\lambda_i^h(f_{hi}) = \lambda_j^h(f_{hj}) = \beta_{\hat i}^h(r_h-f_{hi}) = \beta_{\hat
j}^h(r_h-f_{hj}) \triangleq \eta_h.
\]
\end{lemma}

\textit{Proof:} The replicating response implies $\beta_i^h(f_{hi})^- = \beta_{\hat
i}^h(r_h-f_{hi})^+$. The fact that
\[
f_{hi} = \mathop{\arg\max}_{0\le f \le r_h}\int_0^f \beta_{\hat i}^h(r_h-r) -
\lambda_i^h(r)~dr
\]
implies $\beta_{\hat i}^h(r_h-f_{hi})^+ \ge \lambda_i(f_{hi})^-$ and $\lambda_i(f_{hi})^+
\ge \beta_{\hat i}^h(r_h-f_{hi})^-$. By the same reasoning as used in the proof of
Lemma~\ref{lma:OligVirtualMargin}, it can be shown that $\beta_{\hat i}^h(r_h-f_{hi})^-
\ge \beta_j^h(f_{hj})^-$. Invoking Lemma~\ref{lma:EverywhereComp1}, we have
\be\label{eq:EverywhereComp2Proof1} \beta_i^h(f_{hi})^- = \beta_{\hat i}^h(r_h-f_{hi})^+
\ge \lambda_i^h(f_{hi})^- \ge \lambda_i^h(f_{hi})^+ \ge \beta_{\hat i}^h(r_h-f_{hi})^-
\ge \beta_j^h(f_{hj})^-. \ee By symmetry,
\[
\beta_j^h(f_{hj})^- = \beta_{\hat j}^h(r_h-f_{hj})^+ \ge \lambda_j^h(f_{hj})^- \ge
\lambda_j^h(f_{hj})^+ \ge \beta_{\hat j}^h(r_h-f_{hj})^- \ge \beta_i^h(f_{hi})^-.
\]
Thus, it can be concluded that all terms involved in the above two inequalities must be
equal to each other. So the proof is complete. \qed

\begin{lemma}\label{lma:EverywhereComp3}
Given any (focal) equilibrium with induced link flows $(f_{hi})$ and node total incoming
rates $(r_i)$, for any $h \ne w$ such that $f_{hw}>0$ and $f_{hi}>0$ for a relay $i
\in{\cal O}_h$, it holds that

(i) the actual marginal cost $\lambda_w^h(\cdot)$, $\lambda_i^h(\cdot)$ of $w$ and $i$
forwarding traffic for $h$ are continuous at $f_{hw}$ and $f_{hi}$, respectively;

(ii) the marginal pricing functions $\beta_{\hat i}^h(\cdot)$ of $i$'s virtual competitor
is continuous at $r_h-f_{hi}$;

(iii)
\[
\lambda_i^h(f_{hi}) = \beta_{\hat i}^h(r_h-f_{hi}).
\]
\end{lemma}

\textit{Proof:} First of all, $w$ always uses honest pricing, i.e., $\beta_w^h(\cdot) =
\lambda_w^h(\cdot) = d_{hw}(\cdot)$. So $\lambda_w^h(\cdot)$ and $\beta_w^h(\cdot)$ are
continuous everywhere. For the relay $i$, the inequality~\eqref{eq:EverywhereComp2Proof1}
holds with $j$ being replaced by $w$. Also notice that $\beta_i^h(f_{hi})^- \le
\beta_w^h(f_{hw})^+$. For otherwise, $h$ would be able to strictly reduce its total cost
by shifting an infinitesimal amount of flow from $(h,i)$ to $(h,w)$. However, since
$\beta_w^h(\cdot)$ is continuous, $\beta_w^h(f_{hw})^+ = \beta_w^h(f_{hw})^-$. It follows
that all the inequalities in~\eqref{eq:EverywhereComp2Proof1} must hold with equality. So
the proof is complete. \qed

\begin{lemma}\label{lma:EverywhereComp4}
At an everywhere competitive equilibrium, all the paths with positive flow have equal
marginal cost.
\end{lemma}

\textit{Proof:} Let $s,n_1,n_2,\cdots,n_k,w$ be the nodes on a path ${\cal R}$ with
positive flow at the equilibrium. For simplicity, denote $s$ by $n_0$ and $w$ by
$n_{k+1}$. So $f_{n_i n_{i+1}}
> 0$ for all $i=0,1,\cdots, k$.

At the equilibrium we must have for all $i=1,\cdots,k-1$, \beas
\lambda_{n_i}^{n_{i-1}}(f_{n_{i-1}n_i})^- \stackrel{(a)}{\ge}
d_{n_{i-1}n_i}(f_{n_{i-1}n_i}) + \beta_{n_{i+1}}^{n_i}(f_{n_{i}n_{i+1}})^-
&\stackrel{(b)}{=}&
d_{n_{i-1}n_i}(f_{n_{i-1}n_i}) + \beta_{\hat n_{i+1}}^{n_i}(r_{n_i}-f_{n_{i}n_{i+1}})^+ \\
&\stackrel{(c)}{\ge}& d_{n_{i-1}n_i}(f_{n_{i-1}n_i}) +
\lambda_{n_{i+1}}^{n_i}(f_{n_{i}n_{i+1}})^-. \eeas We have seen inequality $(a)$ in the
proof of Lemma~\ref{lma:EverywhereComp1}. The equality $(b)$ follows from the replicating
response. The inequality $(c)$ is due to the fact that $f_{n_{i}n_{i+1}}$ is the ideal
flow rate to $n_{i+1}$ given $\beta_{\hat n_{i+1}}^{n_i}(\cdot)$. Using the above
relation successively from $i=1$ to $i=k-1$, we obtain \[ \lambda_{n_1}^{s}(f_{sn_1})^-
\ge d_{sn_1}(f_{sn_1})+d_{n_1n_2}(f_{n_1n_2}) + \cdots +
d_{n_{k-2}n_{k-1}}(n_{k-2}n_{k-1}) + \lambda_{n_{k}}^{n_{k-1}}(f_{n_{k-1}n_{k}})^-. \]
Finally notice that $\beta_w^{n_{k}}(\cdot) = d_{n_k w}(\cdot)$, so
\[
\lambda_{n_{k}}^{n_{k-1}}(f_{n_{k-1}n_{k}})^- \ge d_{n_{k-1}n_{k}} +
\beta_{w}^{n_{k}}(f_{n_{k}w})^- = d_{n_{k-1}n_{k}} + d_{n_k w}(f_{n_kw}).
\]
Therefore, the marginal cost of path ${\cal R}$ is upper bounded as
\[
d_{sn_1}(f_{sn_1})+d_{n_1n_2}(f_{n_1n_2}) + \cdots + d_{n_{k-1}n_{k}}(f_{n_{k-1}n_{k}}) +
d_{n_k w}(f_{n_kw}) \le \lambda_{n_1}^{s}(f_{sn_1})^-.
\]

Furthermore, since the equilibrium is everywhere competitive, there exists a node $n_{i}'
\ne n_{i}$ for every $i=1,\cdots,k$ such that $f_{n_{i-1} n_{i}'} > 0$. Using the results
of Lemma~\ref{lma:EverywhereComp2} and \ref{lma:EverywhereComp3}, we can show that
\[
d_{n_{i-1}n_i}(f_{n_{i-1}n_i}) + \lambda_{n_{i+1}}^{n_i}(f_{n_{i}n_{i+1}}) =
d_{n_{i-1}n_i}(f_{n_{i-1}n_i}) + \beta_{\hat n_{i+1}}^{n_i}(r_{n_i}-f_{n_{i}n_{i+1}}) \ge
\lambda_{n_{i}}^{n_{i-1}}(f_{n_{i-1}n_{i}}),
\]
for $i=1,\cdots,k-1$. Applying the above inequality successively from $i=k-1$ to $i=1$,
we have \beas &&d_{sn_1}(f_{sn_1})+d_{n_1n_2}(f_{n_1n_2}) + \cdots +
d_{n_{k-1}n_{k}}(f_{n_{k-1}n_{k}}) + d_{n_k w}(f_{n_kw})\\
&=& d_{sn_1}(f_{sn_1})+d_{n_1n_2}(f_{n_1n_2}) + \cdots +
d_{n_{k-1}n_{k}}(f_{n_{k-1}n_{k}})
+\beta_w^{n_k}(f_{n_kw}) \\
&\ge& d_{sn_1}(f_{sn_1})+d_{n_1n_2}(f_{n_1n_2}) + \cdots + \lambda_{n_k}^{n_{k-1}}\\
&\vdots&\\
&\ge& \lambda_{n_1}^s(f_{sn_1}). \eeas Also by Lemma~\ref{lma:EverywhereComp2},
$\lambda_{n_1}^s(\cdot)$ is continuous at $f_{sn_1}$. Thus, the lower and upper bounds of
the total marginal cost of ${\cal R}$ are both equal to $\lambda_{n_1}^s(f_{sn_1}) =
\eta_s$. Since ${\cal R}$ is chosen arbitrarily, we can conclude that all the paths with
positive flow have the same marginal cost $\eta_s$. \qed

\textit{Proof of Theorem~\ref{thm:EverywhereComp}:} By Lemma~\ref{lma:EverywhereComp4},
at an everywhere competitive equilibrium, every path with positive flow has the same
marginal cost $\eta_s$. To prove that the routing is socially optimal, it remains to show
that any path with zero flow has marginal cost greater than or equal to $\eta_s$. Let
$s,z_1,\cdots,z_m,w$ be the nodes on a zero-flow path ${\cal Z}$. To simplify notation,
write $s$ as $z_0$ and $w$ as $z_{m+1}$. Recall that the flow rate of a path is the
minimum of the flow rates on all its links. So there exist link(s) $(z_i,z_{i+1})$ on
${\cal Z}$ such that $f_{z_i z_{i+1}} = 0$. Next we show that on path ${\cal Z}$,
\be\label{eq:EverywhereCompThmProof1} d_{z_{i-1} z_{i}}(f_{z_{i-1} z_{i}}) + \lambda_{
z_{i+1}}^{z_i}(f_{z_{i} z_{i+1}})^+ \ge \lambda_{ z_{i}}^{z_{i-1}}(f_{z_{i-1} z_{i}})^+,
\ee where $i=1,\cdots,m$.

Notice that for $i=m$,
\[
d_{z_{m-1} z_{m}}(f_{z_{m-1} z_{m}}) + \lambda_{ z_{m+1}}^{z_m}(f_{z_{m} z_{m+1}})^+ =
d_{z_{m-1} z_{m}}(f_{z_{m-1} z_{m}}) + d_{z_{m} w}(f_{z_{m} w}) \ge \lambda_{
z_{m}}^{z_{m-1}}(f_{z_{m-1} z_{m}})^+.
\]
Now consider any $i = 1,\cdots,m-1$. At an everywhere competitive equilibrium, if
$f_{z_{i}z_{i+1}}>0$, then there must exist $z_{i+1}'\ne z_{i+1}$ for which
$f_{z_{i}z_{i+1}'}>0$. Applying Lemma~\ref{lma:EverywhereComp2} or
\ref{lma:EverywhereComp3}, we have
\[
d_{z_{i-1}z_i}(f_{z_{i-1}z_i}) + \lambda_{z_{i+1}}^{z_i}(f_{z_{i}z_{i+1}}) =
d_{z_{i-1}z_i}(f_{z_{i-1}z_i}) + \beta_{\hat z_{i+1}}^{z_i}(r_{z_i}-f_{z_{i}z_{i+1}}) \ge
\lambda_{z_{i}}^{z_{i-1}}(f_{z_{i-1}z_{i}})^+.
\]
This is a special case of~\eqref{eq:EverywhereCompThmProof1} as
$\lambda_{z_{i+1}}^{z_i}(\cdot)$ is continuous at $f_{z_{i}z_{i+1}}$. If
$f_{z_{i}z_{i+1}}=0$, however, we have to consider the following two cases. In the first
case, $r_{z_{i}}>0$, so $z_i$ has at least two offsprings $z_{i+1}'$, $z_{i+1}''$ such
that $f_{z_{i}z_{i+1}'}>0$ and $f_{z_{i}z_{i+1}''}>0$. Assume $z_{i+1}' \ne w$, we
therefore have \beas d_{z_{i-1}z_i}(f_{z_{i-1}z_i}) +
\lambda_{z_{i+1}}^{z_i}(f_{z_{i}z_{i+1}})^+ &=&
d_{z_{i-1}z_i}(0) + \lambda_{z_{i+1}}^{z_i}(0)^+\\
&\stackrel{(a)}{\ge}&
d_{z_{i-1}z_i}(0) + \beta_{\hat z_{i+1}}^{z_i}(r_{z_i})^- \\
&\stackrel{(b)}{\ge}& d_{z_{i-1}z_i}(0) + \beta_{z_{i+1}'}^{z_i}(f_{z_{i}z_{i+1}'})^- \\
&\stackrel{(c)}{=}& d_{z_{i-1}z_i}(0) + \beta_{\hat
z_{i+1}'}^{z_i}(r_{z_i}-f_{z_{i}z_{i+1}'})\ge
\lambda_{z_{i}}^{z_{i-1}}(f_{z_{i-1}z_{i}})^+.\eeas Here, inequality $(a)$ holds because
$f_{z_{i}z_{i+1}}=0$ is the ideal amount of flow to $z_{i+1}$. We applied the same
reasoning as used in the proof of Lemma~\ref{lma:OligVirtualMargin} to get $(b)$.
Inequality $(c)$ follows from the replicating response of $z_{i+1}'$ and from using
Lemma~\ref{lma:EverywhereComp2} or \ref{lma:EverywhereComp3} (depending on whether
$z_{i+1}' = w$ or not). Next we consider the case where $r_{z_{i}}=0$ and consequently
all offsprings of $z_{i}$ adopt honest pricing to $z_{i}$. It follows that
\[
d_{z_{i-1}z_i}(f_{z_{i-1}z_i}) + \lambda_{z_{i+1}}^{z_i}(f_{z_{i}z_{i+1}})^+ =
d_{z_{i-1}z_i}(f_{z_{i-1}z_i}) + \beta_{z_{i+1}}^{z_i}(f_{z_{i}z_{i+1}})^+ \ge
\lambda_{z_{i}}^{z_{i-1}}(f_{z_{i-1}z_{i}})^+.
\]

So far we have proved \eqref{eq:EverywhereCompThmProof1}. Using
\eqref{eq:EverywhereCompThmProof1} we can lower bound the marginal cost of ${\cal Z}$ as
\beas &&d_{z_{0}z_1}(f_{z_{0}z_1}) + \cdots + d_{z_{m-1}z_m}(f_{z_{m-1}z_m})+
d_{z_{m}w}(f_{z_{m}w})\\
&=& d_{z_{0}z_1}(f_{z_{0}z_1}) + \cdots + d_{z_{m-1}z_m}(f_{z_{m-1}z_m})+
\lambda_w^{z_m}(f_{z_{m}w}) \\
&\vdots&\\
&\ge& \lambda_{z_1}^s(f_{sz_1})^+. \eeas

If $f_{sz_1}>0$, by Lemma~\ref{lma:EverywhereComp2}, $\lambda_{z_1}^s(\cdot)$ is
continuous at $f_{sz_1}$ and is equal to $\eta_s$. So we are done. If $f_{sz_1}=0$, $s$
must have two other offsprings $z_1'$, $z_1''$ for which $f_{sz_1'}>0$, $f_{sz_1''}>0$.
Then we can apply the same argument as we used in inequalities $(a)$-$(c)$ to show that
\[
\lambda_{z_1}^s(f_{sz_1})^+ \ge \beta_{\hat z_1'}^s(R_s - f_{sz_1'}) =
\lambda_{z_1'}^s(f_{sz_1'}),
\]
where $\lambda_{z_1'}^s(f_{sz_1'}) = \eta_s$. So we are done.

To summarize, we have shown that at an everywhere competitive equilibrium, every path
with positive flow has the same marginal cost $\eta_s$; moreover, every path with zero
flow has marginal cost greater than or equal to $\eta_s$. Therefore, the routing pattern
of such an equilibrium is socially optimal. \qed

\bibliography{SelfishPricing}

\begin{thebibliography}{10}

\bibitem{paper:IMM05}
O.~Ileri, S.~Mau, and N.~Mandayam, ``Pricing for enabling forwarding in
  self-configuring ad hoc networks,'' {\em IEEE Journal on Selected Areas in
  Communications}, vol.~23, no.~1, pp.~151--162, 2005.

\bibitem{paper:BLV05}
A.~Blanc, Y.~Liu, and A.~Vahdat, ``Designing incentives for peer-to-peer
  routing,'' in {\em Proceedings of IEEE INFOCOM 2005}, vol.~1, Mar. 2005.

\bibitem{paper:CGKO04}
J.~Crowcroft, R.~Gibbens, F.~Kelly, and S.~\"{O}string, ``Modelling incentives
  for collaboration in mobile ad hoc networks,'' {\em Performance Evaluation},
  vol.~57, no.~4, pp.~427--439, 2004.

\bibitem{paper:MQ05}
P.~Marbach and Y.~Qiu, ``Cooperation in wireless ad hoc networks: a
  market-based approach,'' {\em IEEE/ACM Transactions on Networking}, vol.~13,
  pp.~1325--1338, Dec. 2005.

\bibitem{book:BH07}
L.~Buttyan and J.-P. Hubaux, {\em Security and Cooperation in Wireless
  Networks}.
\newblock Cambridge University Press, 2007.

\bibitem{paper:KMT98}
F.~Kelly, A.~Maulloo, and D.~Tan, ``Rate control in communication networks:
  shadow prices, proportional fairness and stability,'' {\em Journal of the
  Operational Research Society}, vol.~49, 1998.

\bibitem{paper:CDR03}
R.~Cole, Y.~Dodis, and T.~Roughgarden, ``Pricing network edges for
  heterogeneous selfish users.,'' in {\em Proceedings of the 35th Annual ACM
  Symposium on Theory of Computing}, pp.~521--530, Jun. 2003.

\bibitem{paper:Rou02}
T.~Roughgarden and E.~Tardos, ``How bad is selfish routing,'' {\em Journal of
  the ACM}, vol.~49, no.~2, pp.~236--259, 2002.

\bibitem{paper:Rou05}
T.~Roughgarden, ``Selfish routing with atomic players,'' in {\em Proceedings of
  the sixteenth annual ACM-SIAM symposium on Discrete algorithms},
  pp.~1184--1185, 2005.

\bibitem{paper:SV03}
J.~Shu and P.~Varaiya, ``Pricing network services,'' in {\em Proceedings of
  IEEE INFOCOM 2003}, vol.~2, Mar. 2003.

\bibitem{paper:HW05}
L.~He and J.~Walrand, ``Pricing differentiated internet services,'' in {\em
  Proceedings of IEEE INFOCOM 2005}, vol.~1, Mar. 2005.

\bibitem{paper:BS02}
T.~Basar and R.~Srikant, ``Revenue-maximizing pricing and capacity expansion in
  a many-users regime,'' in {\em Proceedings of IEEE INFOCOM 2002}, vol.~1,
  2002.

\bibitem{paper:HeW05}
L.~He and J.~Walrand, ``Pricing and revenue sharing strategies for internet
  service providers,'' in {\em Proceedings of IEEE INFOCOM 2005}, vol.~1, Mar.
  2005.

\bibitem{paper:SS05}
S.~Shakkottai and R.~Srikant, ``Economics of network pricing with multiple
  isps,'' in {\em Proceedings of IEEE INFOCOM 2005}, vol.~1, Mar. 2005.

\bibitem{paper:AO06}
D.~Acemoglu and A.~Ozdaglar, ``Competition in parallel-serial networks.'' To
  appear in {\em IEEE Journal of Selected Areas of Communication, Special Issue
  on Non-Cooperative Behavior in Networking}, 2006.

\bibitem{book:Wil93}
R.~Wilson, {\em Nonlinear Pricing}.
\newblock Oxford University Press Inc, USA, 1993.

\bibitem{paper:BW86}
B.~Bernheim and M.~Whinston, ``Menu auctions, resource allocation, and economic
  influence,'' {\em Quarterly Journal of Economics}, vol.~101, no.~1,
  pp.~1--32, 1986.

\bibitem{paper:Rou03}
T.~Roughgarden, ``The price of anarchy is independent of the network
  topology,'' {\em Journal of Computer and System Sciences}, vol.~67, no.~2,
  pp.~341--364, 2003.

\bibitem{book:BG92}
D.~P. Bertsekas and R.~Gallager, {\em Data Networks}.
\newblock Prentice Hall, second~ed., 1992.

\end{thebibliography}
\bibliographystyle{ieeetr}

\end{document}